\documentclass[aps, pra, a4paper, amssymb, reprint, showpacs, nofootinbib]{revtex4-1}

\usepackage{amssymb}
\usepackage{graphicx}
\usepackage{amsmath}
\usepackage{hyperref}

\usepackage{bbold}

\newcommand{\unit}[1]{\text{\ #1}}

\newcommand{\pr}[1]{|#1\rangle\langle#1|}
\newcommand{\prt}[2]{|#1\rangle\langle#2|}
\newcommand{\abs}[1]{\left|#1\right|}
\newcommand{\sabs}[1]{|#1|}

\newcommand{\smean}[1]{\langle #1 \rangle}
\newcommand{\ket}[1]{| #1 \rangle}
\newcommand{\bra}[1]{\langle #1|}
\newcommand{\mL}{\mathcal{L}}
\newcommand{\mD}{\mathcal{D}}
\newcommand{\mE}{\mathcal{E}}

\newcommand{\mO}{\mathcal{O}}
\newcommand{\mP}{\mathcal{P}}
\newcommand{\mQ}{\mathcal{Q}}
\newcommand{\mF}{\mathcal{F}}
\newcommand{\mC}{\mathcal{C}}
\newcommand{\mS}{\mathcal{S}}
\newcommand{\mT}{\mathcal{T}}

\newcommand{\lt}{\tilde\lambda}
\newcommand{\ti}{\tilde}

\newcommand{\re}{\textrm{Re}}
\newcommand{\im}{\textrm{Im}}
\newcommand{\tr}{\textrm{Tr}}
\newcommand{\rmd}{\textrm{d}}
\newcommand{\eff}{\textrm{eff}}
\newcommand{\rom}{\textrm{om}}
\newcommand{\hc}{\textrm{h.c.}}
\newcommand{\etal}{{\it et al.}}

\newcommand{\smf}{{\smash{f}}}

\newcommand{\rss}{\rho_\textrm{ss}^\rom}

\newcommand{\om}{\omega}

\newcommand{\vecc}[1]{\mathbf{#1}}

\newcommand{\figref}[1]{Fig.\,\ref{#1}}
\newcommand{\eeqref}[1]{Eq.\,\eqref{#1}}

\newcommand{\secref}[1]{Sec.\,\ref{#1}}
\newcommand{\appref}[1]{App.\,\ref{#1}}

\begin{document}
\date{\today}
\title{Optomechanical transducers for quantum information processing}

\author{K. Stannigel$^{1,2}$}
\author{P. Rabl$^{1}$}
\author{A. S. S\o{}rensen$^{3}$}
\author{M.~D. Lukin$^{4}$}
\author{P. Zoller$^{1,2}$}

\affiliation{$^1$Institute for Quantum Optics and Quantum Information, 
Austrian Academy of Sciences, 6020
Innsbruck, Austria}
\affiliation{$^2$Institute for Theoretical Physics, University of Innsbruck, 6020 Innsbruck, Austria}
\affiliation{$^3$QUANTOP, Niels Bohr Institute, University of Copenhagen, DK-2100 Copenhagen \O, Denmark}
\affiliation{$^4$Physics Department, Harvard University, Cambridge, Massachusetts 02138, USA}

\begin{abstract}
We discuss  the implementation of optical quantum networks where the interface between stationary and photonic qubits is realized by  optomechanical transducers [K. Stannigel \emph{et al.}, PRL {\bf  105}, 220501 (2010)].  This approach does not rely on the optical properties of the qubit and thereby enables optical quantum communication applications for a wide range of solid-state spin- and charge-based systems. We present an effective description of such networks for many qubits and give a derivation of a state transfer protocol for long-distance quantum communication. We also describe how to mediate local on-chip interactions by means of the optomechanical transducers that can be used for entangling gates. We finally discuss experimental systems for the realization of our proposal.
\end{abstract}


\maketitle

\section{Introduction}

The distribution of quantum information between individual nodes of larger quantum networks is a key  requirement for many quantum information applications \cite{Kimble2008}, in particular for long-distance quantum communication and quantum key distribution protocols \cite{Zeilinger2000}.
In contrast to classical networks, quantum communication channels must allow the distribution of entanglement, which is the essential resource that can be harnessed by further local processing within the nodes. In building high-fidelity quantum channels over long distances, qubits encoded in propagating photons play a unique role, since they provide the only way to transmit quantum information over kilometer distances~\cite{Muller1996,Ursin2007}. 
However, also on a smaller scale rapid progress is made in the design of nano-fabricated photonic circuits for classical and quantum applications \cite{WKB08} and compared to electric circuits
\cite{Bialczak2011,Majer2007} 
or diverse qubit shuttling techniques \cite{Kielpinski2002,Taylor2005}, such photonic circuits could provide a fast and robust alternative for 'on-chip' distribution of entanglement.
Therefore, the development of coherent interfaces between stationary and so-called `flying'  photonic qubits is one of the essential steps in realizing  quantum networks and is commonly included in the list of key criteria for general purpose quantum information processing platforms \cite{DiVincenzo2000}.

\begin{figure}
\includegraphics[width=0.45\textwidth]{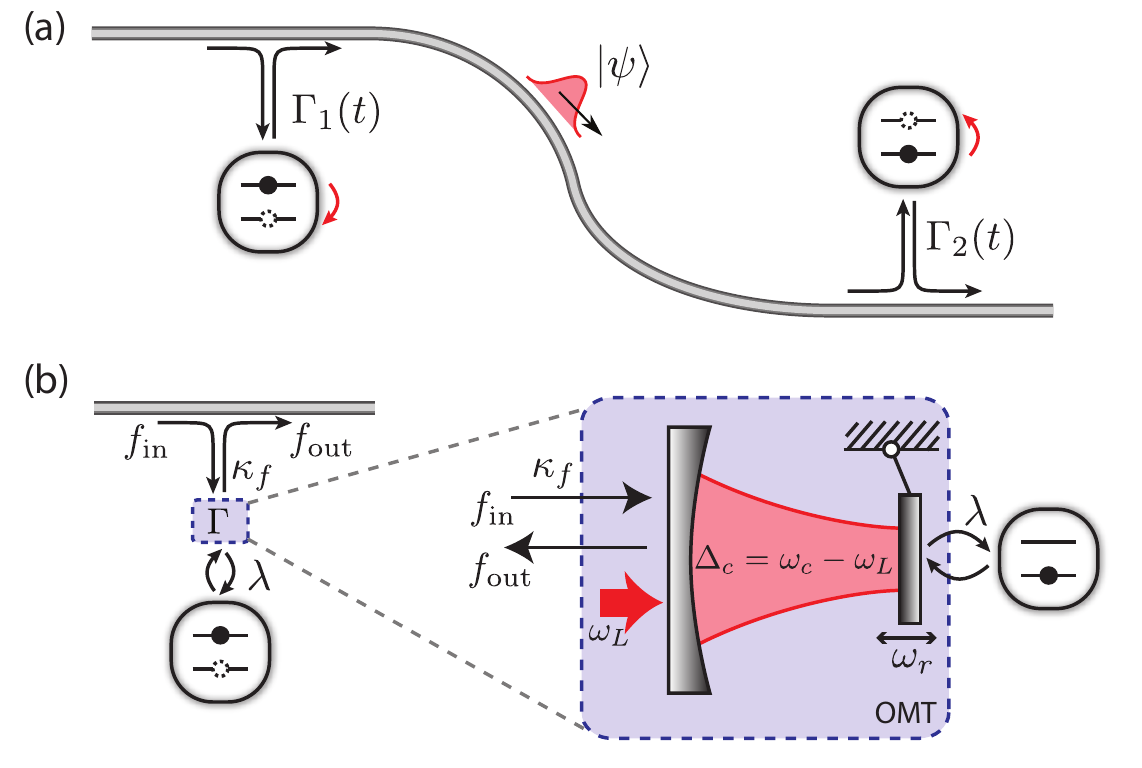}
\caption{(Color online) (a) Elementary quantum network for the realization of state transfer protocols between two nodes via the exchange of photons propagating along an optical fiber. Tunable qubit decay rates $\Gamma_i$ allow for a controlled emission of the photon wave-packet and a perfect reabsorption at the second node. 
(b) Schematic setup for implementing effective qubit-light interfaces based on optomechanical transducers (OMTs): The mechanical resonator mediates a coupling between a solid-state qubit and the driven optical cavity mode. As a result, qubit excitations decay with an effective rate $\Gamma$ into the fiber via the cavity output (see text for details).
}
\label{fig:overview}
\end{figure}

The fundamental operation to be carried out in a quantum network is the transfer of an arbitrary quantum state $\ket{\psi}$ between two nodes according to $\ket{\psi}_1\ket{0}_2\rightarrow \ket{0}_1\ket{\psi}_2$, which together with local operations allows the generation of inter-node entanglement.
Ideas for a physical implementation of this basic building block have first been developed in the context of atomic cavity QED. Here, qubits are stored in spin or hyperfine states of trapped atoms or ions and can be selectively excited to other electronic levels where they couple strongly to light.  As was shown by Cirac \etal \cite{Cirac1997}, this level of control can be used to design a deterministic state transfer protocol based on the tailored emission of a photon, which is subsequently reabsorbed at a second node along an optical fiber (see \figref{fig:overview}(a)).

In parallel to the developments in the field of atomic qubits, substantial progress has  been made on nano-fabricated solid-state qubits, including for example impurity-spins \cite{Morton2011} in various host-materials (such as silicon \cite{Tyryshkin2003} or diamond \cite{WrachtrupJelezko2006,Dutt2007}),  quantum dots 
\cite{Churchill2009,Zwanenburg2009}
or superconducting devices
\cite{Plantenberg2007,DiCarlo2009,Mallet2009,Neeley2010}.
In view of the remarkable level of coherence and control that has already been achieved in such systems,  the challenge is now to identify a suitable interconnection with optical quantum channels also for this larger class of qubits. This is, however, often hindered by lack of coherent optical transitions or the incompatibility with light.  Early suggestions to overcome this problem have been made in the context of hybrid systems, where,  e.g., superconducting devices are coupled to 
close-by atomic \cite{Tian2004,Sorensen2004,Verdu2009}, 
molecular \cite{Rabl2006,Andre2006} 
or spin systems \cite{Schuster2010,Kubo2010,Bushev2011}.  An optical interface could then in principle be realized by using, for example, the atoms as a mediator. 
Recently, an alternative promising route has become available which is fully solid-state based and makes use of the quantized motion of macroscopic mechanical resonators.
In a cavity-optomechanical setting, these resonators can coherently interact with light \cite{KippenbergVahala2008} and their control near the single-quantum level has been demonstrated \cite{WilsonRae2007,*Marquardt2007,*WilsonRae2008,Thompson2008,Groblacher2009,Schliesser2009}. On the other hand, they can also interact with solid-state qubits via magnetic \cite{Mamin2007,Rabl2009,Rabl2010} or electric fields \cite{Armour2002,IrishSchwab2003,Martin2004,LaHaye2009}, and these abilities make them  natural candidates for mediating indirect qubit-photon interactions, thereby realizing the desired qubit-light interface.

In a recent work \cite{Stannigel2010} we proposed the use of such optomechanical transducers (OMTs) for the implementation of optical communication protocols between two solid-state qubits, while related ideas have also been discussed in the context of traveling-wave phonon to photon converters \cite{Safavi-Naeini2011}.
The basic idea of an OMT is illustrated in \figref{fig:overview}(b), where a spin- or charge-based qubit is coupled to the motion of a mechanical resonator via magnetic field gradients 
 or electrostatic interactions. 
The resonator in turn interacts with the field of an optical cavity mode via radiation pressure or optical gradient forces as is currently experimentally explored with various different optomechanical (OM) settings \cite{Thompson2008,Groblacher2009,Schliesser2009,Regal2008,Arcizet2006,Wilson2009}.  As a result, this configuration induces effective interactions between the qubit and photons which do not rely on the optical properties of the qubit system. Therefore, 
this approach in principle enables quantum communication applications for a large range of solid-state qubits.  

The purpose of the present work is two-fold.  In the first and main part of the paper we study long-distance optomechanical quantum networks as introduced in Ref.~\cite{Stannigel2010}. Here, we present a detailed analysis of the effective qubit-fiber interactions and derive a cascaded master equation which describes the effective multiqubit dynamics in such a network. In particular, we describe the implementation of tunable qubit-fiber couplings for quantum state transfer protocols and discuss the role of thermal noise and Stokes-scattering processes. These are the two main noise sources which are typically absent  in analogous atomic settings.   In the second part of the paper we then investigate the case of a single mode optical quantum channel relevant for optical on-chip communication schemes.  In this setting, the qubits are effectively connected by a finite number of mechanical and optical modes and we analyze the performance of entangling operations mediated by this type of network.

The remainder of this paper is structured as follows: In Section \ref{sec:singleNode} we present a detailed description of a single OMT as depicted in \figref{fig:overview}(b) and discuss the resulting effective qubit-fiber interaction together with the relevant noise processes in this system. In \secref{sec:longDistance} we extend the model to an arbitrary number of nodes and derive a cascaded master equation for this multiqubit network. Further, we give a protocol for state-transfer between two nodes and assess its performance. 
In \secref{sec:onchip} we then study an on-chip setting where the individual nodes are coupled by optical resonators with a discrete spectrum and analyze the  prospects for performing entangling gates between the qubits. Finally, we discuss potential experimental realizations of the proposed OMT  in \secref{sec:realizations} and close with concluding remarks in \secref{sec:conclusion}.

\section{An optomechanical interface between stationary qubits and light }
\label{sec:singleNode}

The aim of this work is to develop a theoretical model for the description of optical quantum networks where  the interconversion between stationary and photonic qubits is realized by an OMT at each node. As a first step we will  investigate in this section the dynamics of a single node of such a network and derive a simplified model for the effective qubit-light interactions mediated by the OM device.

\subsection{Model}

We consider a single node of an optical network as schematically shown in~\figref{fig:overview}(b), where the coupling between a qubit  and an optical fiber is mediated by an OMT.  The total Hamiltonian for this  system is
\begin{align}
\label{eq:HsingleNode}
H=H_{\rm node} + H_{\rm fib} + H_\textrm{cav-fib}+H_{\rm env}\,,
\end{align}
where  $H_{\rm node}$ describes the coherent dynamics of the qubit and the OM device, $H_{\rm fib}$ the free evolution of the optical fiber modes, and $H_\textrm{cav-fib}$  the coupling between cavity and fiber. The last term in  \eeqref{eq:HsingleNode}, $H_{\rm env}$, summarizes all additional interactions with the environment leading to decoherence as  specified below.  

For a general discussion we assume complete local control over the qubit, which is taken to be encoded in two states $|0\rangle$ and $|1\rangle$ of an isolated quantum system, and which interacts with the quantized motion of a mechanical resonator, e.g.,
via magnetic field gradients~\cite{Rabl2009} or electrostatic interactions \cite{Armour2002,IrishSchwab2003,Martin2004,LaHaye2009}. The mechanical resonator in turn is coupled to the field of an optical cavity mode via radiation pressure or optical gradient forces. In terms of the usual Pauli operators $\sigma_z=\pr{1}-\pr{0}$ and $\sigma^-=\prt{0}{1}$ for the qubit, and the bosonic operator $b$ for the mechanical mode, the Hamiltonian 
$H_{\rm node}$ can be written in the form  ($\hbar=1$)
\begin{align}
\label{eq:Hnode}
H_{\rm node}&=\frac{\om_q}{2}\sigma_z + \frac{\lambda}{2} (\sigma^- b^\dagger + \sigma^+ b) + H_{\rm om}.
\end{align}
Here, $\omega_q$ is the qubit's energy splitting, $\lambda$ the strength of the qubit-resonator coupling and $H_{\rm om}$ is the Hamiltonian for the coupled OM system. 
We point out that depending on the physical implementation, the Jaynes-Cummings type interaction assumed in \eeqref{eq:Hnode} may emerge only as an effective description of an underlying driven two- or multilevel system. In this case, also $\omega_q$ is an effective parameter which can differ from the bare frequency scales in the system.  
(This will be discussed in more detail in \secref{sec:realizations} where we consider the implementation of the Hamiltonian \eqref{eq:Hnode} for the specific examples of spin and charge qubits.) 

The OM system consisting of a mechanical resonator coupled to a single optical cavity mode is  described by the Hamiltonian 
\begin{align}
\label{eq:Hom}
H_{\rm om}&=\om_r b^\dagger b + \om_{c} c^\dagger c + 
g_{0} c^\dagger c(b+b^\dagger) \\
&\phantom{=} + i\mE(t) e^{-i\om_L t} c^\dagger - i\mE^*(t) e^{i\om_L t} c\,\nonumber,
\end{align}
where $\om_r$  is the mechanical vibration frequency and $c$ is the bosonic annihilation operator for the optical mode of frequency  $\om_c$. The quantity  $g_0=a_0 \frac{\partial \om_c}{\partial x}$ is the single photon OM coupling and corresponds to the optical frequency shift per zero point motion $a_0$.  In the last line of \eeqref{eq:Hom} we have included an additional external laser field of frequency $\omega_L$, which coherently excites the cavity field.  As detailed below, this driving field will allow us to enhance and also to control the OM coupling by adjusting the slowly varying driving strength $\mathcal{E}(t)$. We finally stress that the optical cavity might support additional degenerate modes which we ignored in writing \eeqref{eq:Hom}. As will be discussed below, this is valid as long as the OM interaction does not cause scattering between the degenerate modes.

\subsection{Quantum Langevin equations}
\label{sec:QLEs}

The optical cavity of the OM system is coupled to the modes of an optical fiber, which in the multinode setting considered below will serve as our quantum communication channel. The fiber supports many continua of modes described by bosonic operators $a_{\sigma\om}$, normalized to $[a_{\sigma \omega},a^\dag_{\sigma'\omega'}]=\delta_{\sigma,\sigma'}\delta(\omega-\omega')$, where $\sigma$ labels, e.g., propagation direction or polarization. The free fiber Hamiltonian is then simply given by $H_{\rm fib}=\sum_\sigma\int\rmd\om\, \om\, a_{\sigma\om}^\dag a_{\sigma\om}$. We assume that the OM system is located at position $z=0$ along the fiber to which it is coupled according to
\begin{align*}
H_\textrm{cav-fib}=  i \sum_{\sigma} \int_0^\infty \frac{\rmd\omega}{\sqrt{\pi}}  \left( \sqrt{\kappa_\sigma(\omega)}\, c\, a^\dag_{\sigma\omega} - \hc \right),
\end{align*}
where the coupling constants $\sqrt{\kappa_\sigma(\omega)}$ are defined in this way for later convenience.
For definiteness, we consider the case where the cavity mode is side-coupled to a fiber with left- and right-moving modes of a single relevant polarization ($\sigma=R,L$). In this case, the cavity-fiber coupling is naively estimated to be proportional to the overlap integral
\begin{equation}
\sqrt{\kappa_{R,L}(\omega)} \sim \int_{-\Delta z}^{\Delta z} dz' h(z') e^{\mp i\omega z'/c} \,,
\end{equation}
where $h(z)$ is the mode function of the evanescent cavity field over the relevant interaction region of length $2\Delta z$ (see, e.g., Refs.\,\cite{Gorodetsky1999,Min2007} for a detailed analysis of the coupling to a whispering-gallery mode cavity).
For $\Delta z \gtrsim \lambda_c$, where $\lambda_c$ is the vacuum cavity wavelength, we can roughly distinguish between two cases.  For a standing-wave cavity mode  $h(z)\sim \cos(\omega_c z/c)$  and photons from the cavity will be emitted into both directions ($\kappa_{R}\approx \kappa_{L}$), while for a running-wave mode such as $h(z)\sim e^{ i\omega_c z/c}$, a preferred emission into a specific direction can be achieved  (in this case $\kappa_{R}\gg \kappa_{L}$). Although in principle both configurations could be relevant for different quantum communication applications, we will for concreteness focus in the following on the case where $c$ describes a circulating running-wave cavity mode that couples to the fiber according to
\begin{align}
\label{eq:HcavfibSingleNode}
H_\textrm{cav-fib} &=  i   \sqrt{2\kappa_f}  \left( c f_R^\dag(z=0) - c^\dag f_R(z=0) \right),
\end{align}
where $f_R(z)=  \frac{1}{\sqrt{2\pi}}\int_0^\infty \rmd\om\, a_{R \omega} e^{i \omega z/c}$ is the right-propagating fiber-field. The coupling of the cavity to this continuum of modes will lead to an irreversible decay of the cavity field at the rate $\kappa_f\equiv \kappa_R(\omega_c) $.

To proceed, we eliminate the time-dependence in \eeqref{eq:Hom} by transforming all photonic operators to a frame rotating at the laser frequency. This amounts to replacing $H_{\rm om}$ by
\begin{align}
\label{eq:HomBar}
\bar H_{\rm om}&=\om_r b^\dagger b + \Delta_c c^\dagger c + g_{0} c^\dagger c(b+b^\dagger)  \\
&\phantom{=}+i\left(\mE(t) c^\dagger -\mE^*(t) c\right)\,\nonumber,
\end{align}
and $H_{\rm fib}$ by $\bar H_{\rm fib}=\sum_\sigma\int\rmd\om\, \Delta_\om\, a_{\sigma\om}^\dag a_{\sigma\om}$\,,
where $\Delta_c=\om_c-\om_L$ and $\Delta_\om=\om-\om_L$ are the detunings of the cavity and fiber modes from the laser frequency, respectively. We then  use a standard Born-Markov approximation to eliminate the fiber modes \cite{QuantumNoise} and describe the resulting dissipative dynamics in terms of a quantum Langevin equation (QLE) for the cavity field, 
\begin{align}
\label{eq:singleNodeQLE1}
\dot c &\approx -i[c,H_{\rm node}] - \kappa c  - \sqrt{2\kappa_f} f_{\rm in}(t) - \sqrt{2\kappa_0} f_{0}(t) \,.
\end{align}
Here, we have introduced a total decay rate $\kappa=\kappa_f+\kappa_0$, where $\kappa_0$ accounts for additional intrinsic losses of the optical cavity mode and $f_0(t)$ is the associated noise operator.  We have further defined operators $f_{\rm in}(t):= f_R(0^-,t)$ and $f_{\rm out}(t):= f_R(0^+,t)$ which obey the input-output relation
\begin{equation}
\label{eq:cavInOut}
f_{\rm out}(t)\approx f_{\rm in}(t)+\sqrt{2\kappa_f}\, c(t)\,,
\end{equation}
and we specify vacuum white noise statistics for the fiber input, i.e., $[f_{\rm in}(t),f^\dagger_{\rm in}(t^\prime)]=\delta(t-t^\prime)$ and  $\smean{f_{\rm in}(t)f^\dagger_{\rm in}(t^\prime)}=\delta(t-t^\prime)$, as well as for $f_0(t)$. 
In addition, the mechanical resonance generally has a finite intrinsic width and is connected via its support to a thermal bath of temperature $T$. This is commonly modeled by the QLE
\begin{align}
\label{eq:singleNodeQLE2}
\dot b &= -i[b,H_{\rm node}] - \frac{\gamma_m}{2}b -\sqrt{\gamma_m}\xi(t)\,,
\end{align}
where $\gamma_m/2$ is the decay rate of the mechanical amplitude.
In contrast to the optical fields, the mechanical noise is characterized by a thermal white noise operator $\xi(t)$  with statistics 
$\smean{\xi(t)\xi^\dagger(t^\prime)}=(N_m+1)\delta(t-t^\prime)$ and 
$[\xi(t),\xi^\dagger(t^\prime)]=\delta(t-t^\prime)$. Here, $N_m$ is given by the Bose occupation number for a mode of frequency $\om_r$. In the high-temperature case, where $k_B T \gg \hbar\om_r$ and hence $N_m\gg1$, the relevant thermal decoherence rate will turn out to be $\gamma_m N_{m}\approx\frac{k_B T}{\hbar Q}$, where $Q=\om_r/\gamma_m$ is the quality factor of the mechanical resonance and typically $Q\gg1$. Below, we will often take the limit $\gamma_m\rightarrow 0$ and whenever doing so it is understood that we keep the decoherence rate $\gamma_m N_{m}$ constant.

\subsection{Linearized equations of motion}
\label{sec:singleNodeLinearization}

In typical experiments the OM coupling $g_0$ is too small to allow coherent interactions between the mechanical system and individual cavity photons.
Therefore, to achieve an appreciable and also tunable coupling we consider the case of a strongly driven OM system where the effective coupling between mechanics and light is amplified by the coherent field amplitude inside the cavity.  

Starting from the QLEs\,\eqref{eq:singleNodeQLE1},\eqref{eq:singleNodeQLE2}, we perform a unitary transformation $c\rightarrow c+\alpha$ and $b\rightarrow b + \beta$ such that the c-numbers $\alpha$, $\beta$ describe the classical mean values of the modes and the new operators $c$ and $b$ represent quantum fluctuations around them. We require any classical (that is c-number) contributions to the transformed QLEs to vanish, which yields the following system for the classical response:
\begin{subequations}
\label{eq:classicalResponse}
\begin{align}
\dot \alpha &= -(i\Delta_{c}+\kappa)\alpha -i g_0 \, (\beta+\beta^*)\,\alpha +\mE(t)\,, \\
\dot \beta  &= -(i\om_r+\gamma_m/2) \beta -ig_0 \abs{\alpha}^2\,.
\end{align}
\end{subequations}
For $\mE(t)=\textrm{const}$ and $\gamma_m \rightarrow 0$ the steady state solution is $\beta=-g_0\sabs{\alpha}^2/\om_r$ with $\alpha$ determined from 
\begin{align}
\label{eq:adiabaticE}
\alpha=\frac{\mE }{i\Delta_c + \kappa - 2 i \, g_0^2 \abs{\alpha}^2/\om_r}\,.
\end{align}
This relation is still approximately valid if $\mE(t)$ is slowly varying compared to the characteristic response time. This means, e.g., $\dot\mE(t)/\mE(t)\ll\kappa,\Delta_c,\om_r$ for the case $\abs{\alpha} g_0\ll\Delta_c,\om_r$ which is of relevance below. 
Within these limits, a desired temporal profile $\alpha(t)$ can be directly related to the applied laser power and phase by means of \eeqref{eq:adiabaticE}. For more rapid variations the system \eqref{eq:classicalResponse} has to be integrated exactly to capture all retardation effects. 

In the strong driving regime where $\sabs{\alpha(t)}^2\gg 1$ the OM coupling of \eeqref{eq:Hom} can be linearized: After going to the displaced operators in the QLEs\,\eqref{eq:singleNodeQLE1},\eqref{eq:singleNodeQLE2} and eliminating the classical forces, we drop terms of order $g_0$ relative to those of order $g_0\abs{\alpha}$ and $g_0\abs{\alpha}^2$. The result of this procedure is equivalent to replacing $\bar H_{\rm om}$ by the linearized OM Hamiltonian
\begin{align}
\label{eq:Homlin}
H_{\rm om}^{{\rm lin}}(t)= \om_r b^\dagger b + \tilde\Delta_{c}(t) c^\dagger c 
+(G(t) c^\dagger + G^*(t) c)(b+b^\dagger)\,.
\end{align}
Here, we have introduced the laser-enhanced OM coupling $G(t)=g_0 \alpha(t)$, which describes interconversion between phonons and photons, as well as the renormalized cavity detuning 
$\tilde \Delta_{c}(t)=\Delta_c+g_0(\beta+\beta^*)=\Delta_c - 2\abs{G(t)}^2/\om_r$.
Although the latter shift is small in the regime of interest identified below, one has to take it into account when varying $G(t)$. We will drop the tilde in what follows and also note that we have ignored a contribution $\lambda/2(\beta(t) \sigma^+ + \hc)$ to $H_{\rm node}$, which can be compensated by local fields acting on the qubit.
Finally, if the cavity supports many degenerate modes $c_\mu$ that couple to the mechanical resonator according to $g_0 \sum_\mu c_\mu^\dagger c_\mu(b+b^\dagger)$, then only the OM coupling to the driven mode is enhanced. The others may then be neglected due to $g_0\ll \abs{G}$, which justifies our single-mode treatment. However, this argument breaks down if there is phonon-assisted scattering between the modes, i.e., for an interaction of the form $g_0 \sum_{\mu,\nu} c_\mu^\dagger c_\nu(b+b^\dagger)$.

In summary, the OM system can be described by an enhanced, linear coupling $\abs{G}\gg g_0$ in $H_{\rm om}^{{\rm lin}}$ between the resonator and an undriven cavity mode. The qubit dynamics is determined by the Heisenberg equations of motion
\begin{subequations}
\label{eq:qubitQLE}
\begin{align}
\dot \sigma^-      &= -i \omega_q \sigma^-  + i\frac{\lambda}{2} \sigma_z b\,,\\
\dot \sigma_z     & =- i\lambda (\sigma^+ b - b^\dag\sigma^- )\,,
\end{align}
\end{subequations}
and the linearized OM dynamics (QLEs\,\eqref{eq:singleNodeQLE1},\eqref{eq:singleNodeQLE2} with $H_{\rm om}\rightarrow H_{\rm om}^{\rm lin}$) can be written compactly by introducing the vectors  $\vecc{v}=(b,c,b^\dagger,c^\dagger)^T$ and $\vecc{S}=(\sigma^-,0,-\sigma^+,0)^T$:
\begin{equation}
\label{eq:linSingleNodeQLE}
\dot {\vecc{v}}(t) = -M \vecc{v}(t)  -i \frac{\lambda}{2}\vecc{S}(t) - \vecc{{N}}(t)\,.
\end{equation}
Here, the drift matrix $M$ describes the response of the OM system and is explicitly given by
\begin{align}
\label{eq:OMmatrix}
M=
i \begin{pmatrix}
\om_r -i\frac{\gamma_m}{2}      &  G^*        & 0       &   \zeta G \\
G           & \Delta_c -i\kappa   & \zeta G & 0        \\
0           & -  \zeta G^* & - \om_r -i\frac{\gamma_m}{2}& - G      \\
- \zeta G^* & 0           & -G^*    & -\Delta_c -i\kappa 
\end{pmatrix}\,,
\end{align}
where $\zeta=1$ gives the full linearized OM coupling and $\zeta=0$ corresponds to the often-applied rotating wave approximation (RWA), which is valid for $\kappa,\abs{G}\ll\Delta_c\approx\om_r$ and renders the coherent part of the dynamics excitation-number- or energy-conserving. The noise sources driving the OM system are summarized in $\vecc{N}(t)=\sqrt{2\kappa_f}\vecc{I}(t) +\vecc{R}(t)$, where $\vecc{I}(t)=(0,f_{\rm in},0,f^\dagger_{\rm in})^T$ represents the fiber input-field and the intrinsic noise is contained in $\vecc{R}(t)=(\sqrt{\gamma_m}\xi,\sqrt{2\kappa_0}f_{0},\sqrt{\gamma_m}\xi^\dagger,\sqrt{2\kappa_0}f^\dagger_{0})^T$. The statistics of these noise vectors follow directly from the definitions in the previous subsection and can also be found in \appref{sec:appLongFiber}. 
We finally note that the system described by $M$ generally exhibits self-oscillations for blue detuning $\Delta_c<0$ as well as bi-stability for red detuning $\Delta_c>0$ and strong coupling $\abs{G}^2>(\kappa^2+\Delta_c^2)\om_r/4\Delta_c$ (see, e.g., Ref.\,\cite{Genes2009}). However, in this work we will only consider parameters where these phenomena do not occur.

\subsection{Effective qubit-light interface}
\label{sec:elimination}

\begin{figure*}
\includegraphics[width=0.9\textwidth]{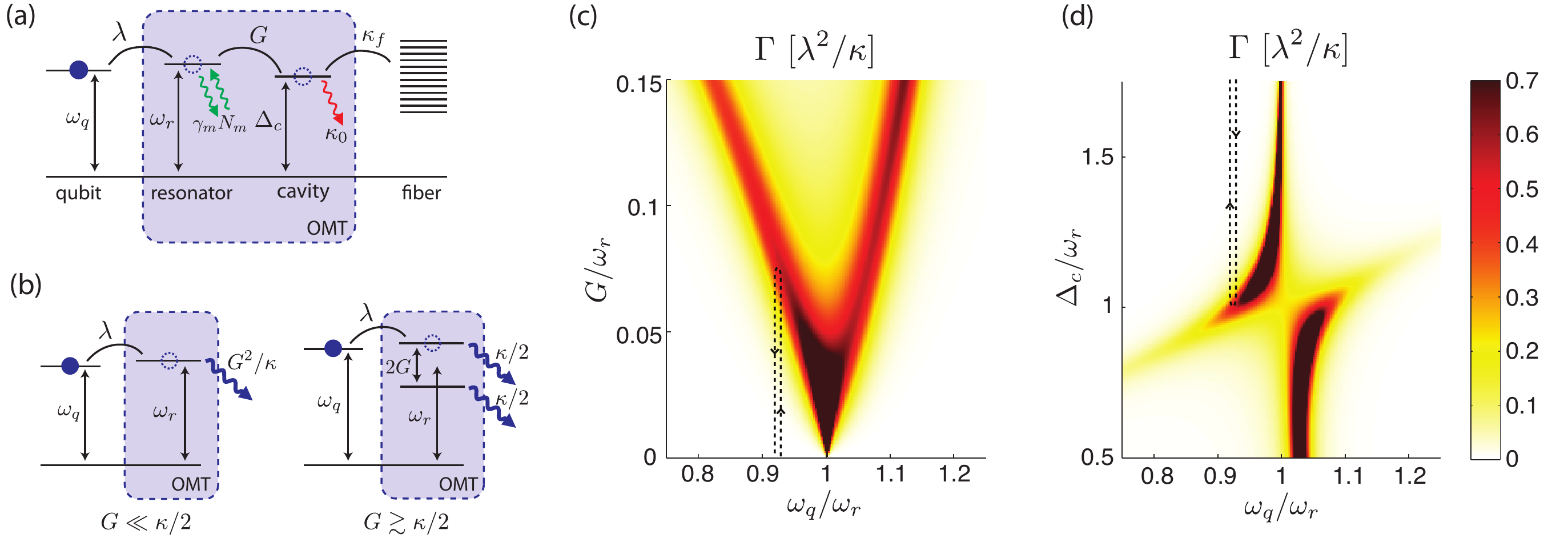}
\caption{(Color online)  
(a) Level scheme illustrating the decay of a qubit excitation into the fiber (only processes corresponding to $\zeta=0$ are show). The dashed box represents the OMT and wavy arrows indicate mechanical diffusion and intrinsic cavity loss. 
(b) Level structure of the OMT as seen by the qubit. Wavy arrows indicate decay into the optical fiber at $\gamma_{\rm op}\approx{\rm min}\{G^2/\kappa,\kappa/2\}$ in the regimes of weak (left) and strong (right) OM coupling.
(c) Effective qubit-decay $\Gamma$ as a function of the linear OM coupling $G$ and the qubit frequency $\om_q$ as given by \eeqref{eq:effCoeffs}. We have used $\kappa=0.05\om_r$ and $\gamma_m=0$, and $\Delta_c=\om_r$ is satisfied at $G=3\kappa/2$. 
(d) $\Gamma$ as a function of cavity detuning and qubit frequency, where $G=3\kappa/2$ at $\Delta_c=\om_r$. The dashed lines in (c),(d) indicate possible paths for tuning $\Gamma$ between $\sim0$ and $\sim\lambda^2/2\kappa$ (see text).
}
\label{fig:singleNode}
\end{figure*}

The dynamics of a single node as described by Eqs.\,\eqref{eq:qubitQLE} and \eqref{eq:linSingleNodeQLE} allows for a coherent conversion of a qubit excitation into a photon propagating in the optical fiber, as is indicated in \figref{fig:singleNode}(a):
First, the excitation is transferred to the mechanical resonator at a rate $\sim\lambda$, from where it is then up-converted by the excitation-number conserving terms of the OM interaction into a cavity photon with a rate $\sim G$, and subsequently emitted into the fiber at a rate $\sim\kappa_f$.  At the same time, this transfer is affected by decoherence in form of mechanical dissipation and photon loss, 
thereby degrading the fidelity of the interface under realistic conditions. 
Our goal is now to derive a simplified model for the interface which captures all of these aspects in an effective description for the qubit-fiber coupling only.

To proceed, let us first look at the dynamics of the linearized OMT, which is fully characterized by the drift matrix $M$ defined in \eeqref{eq:OMmatrix} (see also Ref.\,\cite{Dobrindt2008} for an extensive discussion). It has two independent eigenvalues which we identify with $i\om_\pm+\gamma_\pm$, where $\om_\pm>0$ are the frequencies and $\gamma_\pm$ the decay rates of the associated OM normal modes (the remaining eigenvalues are simply complex conjugates). 
We take $G$ to be real and positive for simplicity and focus on the case $\Delta_c\approx \om_r$ relevant for resonant excitation transfer, where we can distinguish between two regimes:  For weak OM coupling, $G<\kappa/2$, the resonator and cavity modes retain their respective characters and we have, e.g., $\omega_-\approx \omega_r$, $\omega_+\approx \Delta_c$. However, due to a small admixing between the modes the optical cavity provides an additional decay channel for the mechanical system and can lead to significant modifications of the mechanical damping rate. In the limit $G\ll \kappa$ it is given by $\gamma_-= \gamma_m/2 + G^2/\kappa$  (see left panel of \figref{fig:singleNode}(b).  In the opposite regime of strong OM coupling, $G>\kappa/2$, resonator and cavity hybridize and the OM system exhibits a normal mode splitting ($\omega_\pm \approx \omega_r \pm G$).  This hybridization also implies that both modes decay with the same rate $\gamma_{\pm}\approx\kappa/2$ (right panel of \figref{fig:singleNode}(b)). In more general terms we can introduce an OM damping rate $\gamma_{\rm op}$, which for arbitrary parameters is defined as $\gamma_{\rm op}={\rm min}\{\gamma_+,\gamma_-\}$ and reproduces $\gamma_{\rm op}\approx G^2/\kappa$ and $ \gamma_{\rm op}\approx\kappa/2$ in the two respective limits. As $\gamma_{\rm op}$ is the rate at which mechanical excitations are converted into  traveling photons, it sets a maximal transfer rate for the OMT. We expect $\gamma_{\rm op}\gg\gamma_m/2$ for typical parameters, which justifies taking limit $\gamma_m\rightarrow 0$.

We now consider the experimentally relevant regime where the qubit-resonator coupling $\lambda$ is small compared either to $\gamma_{\rm op}$ or to the detuning from the nearest normal mode, that is, the regime where $\lambda\ll{\rm max}(\gamma_{\rm op},{\rm min}(\abs{\om_{\pm}-\om_q}))$. In this limit, the dynamics of the OM system is fast compared to the qubit-resonator coupling and can hence be adiabatically eliminated, yielding the desired effective description of the OMT. 
To this end, we formally integrate \eeqref{eq:linSingleNodeQLE} to obtain
\begin{equation}
\vecc{v}(t)=\vecc{v}_{\rm free}(t) - i\frac{\lambda}{2} \int_{-\infty}^t\rmd s\,e^{-M(t-s)}  {\bf S}(s)\,,
\end{equation}
where we have neglected transients and $\vecc{v}_{\rm free}(t)=-\int_{-\infty}^t\rmd s\,e^{-M(t-s)}\vecc{N}(s)$ is the OM steady state solution in the absence of the qubit (see \eeqref{eq:linSingleNodeQLE} for $\lambda\rightarrow0$). We may now in a Born-Markov approximation and to lowest order in $\lambda$ replace the qubit operators by their free evolution, i.e., $\sigma^-(s)=\sigma^-(t) e^{i\omega_q (t-s)}$ and re-insert the result into \eeqref{eq:qubitQLE}. By keeping only the resonant terms, we obtain effective QLEs for the qubit operators,
\begin{subequations}
\label{eq:effQubitQLEs}
\begin{align}
\dot\sigma^- &= -\left[i(\om_q+\Delta_0) + \frac{\Gamma}{2} \right] \sigma^- - \sqrt{\Gamma} \sigma_z F_{\rm in}(t)\,,\\
\dot\sigma_z &= -\Gamma (\mathbb{1}+\sigma_z)  + \sqrt{4\Gamma}\left( \sigma^+ F_{\rm in}(t) +F^\dag_{\rm in}(t)  \sigma^-  \right)\,,
\end{align}
\end{subequations}
with noise operators $F_{\rm in}(t)=-i \sqrt{\lambda^2/4\Gamma}\, b_{\rm free}(t)$. From \eeqref{eq:cavInOut} we further obtain an  effective input-output relation for the fiber field,
\begin{align}
\label{eq:effInOut}
f_{\rm out}(t)
&\approx f_{\rm in}(t) + \sqrt{2\kappa_{\smf}} \,c_{{\rm free}}(t) - \sqrt{\eta\Gamma} e^{i\phi}\sigma^-(t)\,.
\end{align}
In Eqs.\,\eqref{eq:effQubitQLEs}-\eqref{eq:effInOut} we have introduced the effective decay rate $\Gamma$ and frequency shift $\Delta_0$ by the relations
\begin{align}
\label{eq:effCoeffs}
\Gamma=\frac{\lambda^2}{2} \re\{A_{11}(\om_q)\}\,, \qquad
\Delta_0=\frac{\lambda^2}{4} \im\{A_{11}(\om_q)\}\,,
\end{align}
where $A(\om)=(M-i\om\mathbb{1})^{-1}$ is the OM response matrix.
Further, $\eta=\kappa_f/\kappa$ is the branching ratio resulting from intrinsic cavity decay and $\phi={\rm arg}\{i A_{21}(\om_q)\}$. Note that in writing \eeqref{eq:effInOut} we have neglected small corrections related to finite $\gamma_m$, as well as non-RWA corrections and counter-rotating terms $\propto \sigma^+(t)$.

The effective equations of motion 
\eqref{eq:effQubitQLEs} 
together with the input-output relation 
\eqref{eq:effInOut} 
are familiar from atom-light interactions \cite{QuantumNoise} and describe a two level system which is directly coupled to an optical fiber with an effective decay rate $\Gamma$. The OM system which mediates this interaction has disappeared from the dynamics, but  still determines $\Gamma$, $\Delta_0$, $\phi$, which can thus be controlled by appropriately adjusting  OM parameters. This can be seen in more detail from the expression for the decay rate, 
\begin{equation}\label{eq:decayRate}
\Gamma\approx \frac{\lambda^2 G^2 \kappa/2}{(G^2 +(\Delta_c-\om_q)(\omega_q-\omega_r))^2+ \kappa^2(\omega_q-\omega_r)^2}\, ,
\end{equation}
where we have assumed $\zeta=0$ and $\gamma_m\rightarrow0$.
A plot of the full expression from \eeqref{eq:effCoeffs} is given in \figref{fig:singleNode}(c) for the case $\om_r\approx\Delta_c$  and one clearly observes the underlying mode structure of the OM system as discussed above. 
The decay $\Gamma$ is suppressed if the qubit is detuned from the OM eigenmodes, while under resonance conditions, $\omega_q\approx\omega_\pm$, it scales as $\Gamma\sim  \lambda^2/\gamma_{\rm op}$, and we obtain $\Gamma\approx \lambda^2 \kappa/2G^2$ and $\Gamma\approx \lambda^2/2\kappa$ in the weak and strong coupling regime, respectively. 

As we will discuss in more detail in \secref{sec:stateTransfer} below, it is essential for the implementation of deterministic quantum state transfer protocols within quantum networks to have control over the shape of the emitted photon, which can be accomplished by considering a time-dependent decay rate $\Gamma\rightarrow\Gamma(t)$. To achieve this, we may make use of the fact that in principle any of the OM parameters entering $\Gamma$ could be varied in an experiment, that is, we consider $\Gamma(t)=\Gamma(G(t),\Delta_c(t),\om_q(t),...)$. The effective Markovian description of the qubit dynamics and the algebraic relation in \eeqref{eq:effCoeffs} remain valid as long as the OM parameters are tuned slowly relative to the time-scale of the effective resonator decay $1/\gamma_{\rm op}$,
such that $\dot G/G \ll \gamma_{\rm op}$, etc. In this sense we may adiabatically adjust the effective qubit decay rate and hence acquire control over the generated signal photon.

\emph{Examples}. As a specific example we consider tuning the OM coupling along a path as indicated by the dashed line in \figref{fig:singleNode}(c). Here, $G(t)$ varies between 0 and $\sim 3\kappa/2$ while the qubit frequency is fixed at $\om_q\approx\om_r-3\kappa/2$ and $\om_r\approx\Delta_c$. This corresponds to tuning the lower 
normal mode in and out of resonance with the qubit, such that $\Gamma$ can be varied within a certain range $[\Gamma_{\rm min},\Gamma_{\rm max}]$. For $G \rightarrow 0$  the qubit is far detuned by $\delta=\sabs{\om_q-\om_r}$ from the resonator mode to which it couples and therefore, the effective decay rate is suppressed according to $\Gamma\sim\lambda^2\gamma_{\rm op}/2\delta^2\approx\lambda^2 G^2/2\kappa\delta^2$, which realizes $\Gamma_{\rm min}\rightarrow 0$. A residual decay is still provided by mechanical damping $\Gamma_{\rm min}\approx\lambda^2\gamma_m/4\delta^2$, which, however, is negligible for the parameters considered below. By increasing $G$ up to $\sim3\kappa/2$ we reach the opposite limit of strong OM coupling, where the lower OM normal mode is in resonance with the qubit and a maximal decay of  $\Gamma_{\rm max}=\lambda^2/2\kappa$ is achieved. Within these limits, a desired time-dependence of the effective decay $\Gamma(t)$ can thus directly be translated into a profile $G(t)$ via \eeqref{eq:effCoeffs}.

Similarly, we can vary the cavity frequency along the dashed line in \figref{fig:singleNode}(d), where we chose $\om_q\approx\om_r-3\kappa/2$ and it is assumed that the driving laser is at constant power. This means that $G$ varies according to \eeqref{eq:adiabaticE}, which has been included in the plot. We choose $G=3\kappa/2$ at $\Delta_c=\om_r$, such that $\Gamma_{\rm max}$ for this path is the same as for the previous one. Starting from this point, the effective decay can be turned off by increasing the cavity detuning as far as needed, with a residual decay due to the resonator as before.

\subsection{Noise and imperfections}
\label{sec:singleNodeImperfections}

As already mentioned above, the OMT does not only mediate the desired coherent interactions between the qubit and the optical channel, but under realistic conditions also adds noise. In the effective QLEs\,\eqref{eq:effQubitQLEs} this noise appears first of all in form of the operator $F_{\rm in}(t)$, which generally represents an effective  non-vacuum input that heats the qubit. Second, noise processes also affect the out-field $f_{\rm out}(t)$ given in \eeqref{eq:effInOut}  where on the one hand the term $c_{\rm free}(t)$ represents excess photons which are emitted into the fiber independently of the qubit state and on the other hand $\eta<1$ accounts for loss of qubit excitations through the cavity's intrinsic decay channels.

Let us first look at the noise operator $F_{\rm in}(t)$ which appears in the QLEs \eqref{eq:effQubitQLEs} for the qubit. From its definition  in terms of the OM drift matrix $M$ we see that two-time correlations of $F_{\rm in}(t)$ decay with $\gamma_{\rm op}\gg \Gamma$. Therefore, on the scale of the effective qubit dynamics, it can be considered as nearly white noise and is characterized by an effective occupation number
\begin{align}
N_0 = 2\, \re\,\int_0^\infty \rmd t\, \smean{F^\dag_{\rm in}(t)F_{\rm in}(0)}e^{-i\om_q t}\,.
\end{align}
This expression can be readily evaluated in Fourier space as shown in \appref{sec:appLongFiber} and for $\gamma_m \rightarrow 0$ we obtain
\begin{align}
\label{eq:N0}
N_{0}&\approx \frac{\gamma_m N_m}{2\kappa}\frac{\kappa^2+ (\Delta_{c}-\om_q)^2}{G^2}
 + \frac{\kappa^2+(\Delta_c-\om_q)^2}{4\Delta_c\om_q} \, .
\end{align}
The first contribution (given for $\zeta=0$) is due to the thermal noise of the resonator's environment and roughly scales as $\gamma_m N_m/\gamma_{\rm op}$. 
The second contribution to $N_0$ describes heating due to non-energy-conserving terms in the OM coupling, $\sim G c^\dag b^\dag$, and accounts for photons scattered from the strong driving beam into the cavity while simultaneously creating a mechanical excitation. 
We see that for a qubit on resonance with an OM normal mode ($\om_q\approx\Delta_c\pm G$) the conditions for small thermal noise are $\gamma_m N_m\ll\gamma_{\rm op}$, while to reduce the Stokes scattering events we need $\kappa,G\ll\Delta_c,\om_r$. The latter conditions describe the so-called resolved side-band regime and therefore, we conclude that the conditions for a low noise OMT are equivalent to the requirements for OM ground state cooling \cite{WilsonRae2007,*Marquardt2007,*WilsonRae2008}.

The second type of noise is the contamination of the cavity output with noise photons originating from up-converted thermal noise or Stokes scattering events, as described by the term $\propto c_{\rm free}(t)$ in \eeqref{eq:effInOut}. 
To analyze the effect of these excess photons let us in a first step imagine that at time $t=0$ the qubit is prepared in the excited state and that we record the number of photons emitted into the fiber during a time-interval $T\sim1/\Gamma$. The observed photon number is given by $N(T)=\int_0^T \, \rmd t\, \smean{f_{\rm out}^\dagger f_{\rm out}}(t)$ and for simplicity, we assume that the qubit operator evolves according to $\sigma^-(t)=e^{-(\Gamma/2+i\om_q)t}\sigma^-(0)$, while being uncorrelated with $c_{\rm free}(t)$. We then obtain $N(T)=1-e^{-\Gamma T} +N_{\rm ex}(T)$ with the number of excess photons given by
\begin{align}
N_{\rm ex}(T)=2\kappa_f \int_0^T\rmd t\, \smean{c_{\rm free}^\dagger c_{\rm free}}(t)\,,
\end{align}
which simply yields $N_{\rm ex}(T)=2\kappa_f T\, \smean{c_{\rm free}^\dagger c_{\rm free}}(0)$ for time-independent OM parameters. For $T=\Gamma^{-1}$, $\kappa_0=0$, $\gamma_m\rightarrow 0$, and $\om_r=\Delta_c$ this quantity can be estimated
\footnote{The steady state cavity occupation can be expressed as ${\langle} c^\dagger c{\rangle}_{\rm free}=\int\frac{\rmd\om}{2\pi} \mathcal{C}^{11}_{22}(\om)$, with the matrix $\mathcal{C}^{11}(\om)$ defined in \appref{sec:appLongFiber}. Neglecting non-RWA corrections to the thermal contribution, we obtain 
${\langle} c^\dagger c{\rangle}_{\rm free}\approx\int\frac{\rmd\om}{2\pi}\left(\gamma_m N_m \sabs{A_{21}(\om)}^2 + 2\kappa \sabs{A_{24}(\om)}^2 \right)$ which we evaluate to leading order in $G^2/\om_r^2$
\cite{*[{See integral 3.112.1 in }] [{ and note that there, a factor $(-1)^{n+1}$ is missing (K. Jaehne, Ph.D. thesis, Innsbruck, 2009).}] Gradshteyn2007}.
}
to be 
$N_{\rm ex}(\Gamma^{-1}) \approx \gamma_m N_m/\Gamma + (\kappa/\Gamma)( G^2/\om_r^2)$, and for a qubit resonant with an OM mode we may use $\Gamma\sim \lambda^2/\gamma_{\rm op}$ to obtain the scaling
\begin{align}
\label{eq:Nex}
N_{\rm ex}(\Gamma^{-1}) \sim \frac{\gamma_{\rm op}^2}{\lambda^2} \left( \frac{\gamma_mN_m}{\gamma_{\rm op}}+ \frac{\kappa}{\gamma_{\rm op}}\frac{G^2}{\om_r^2}\right)\,.
\end{align}
While the terms in parentheses are small under the low noise conditions identified above, the prefactor is  much larger than one due to the condition $\gamma_{\rm op}\gg\lambda$ underpinning the adiabatic elimination. The number of excess photons may thus exceed one and make the signal photon difficult to detect in general. However, this estimate did not take into account the fact that the qubit emits in a bandwidth $\Gamma$ around $\om_q$, while the OM noise floor is spread over a bandwidth $\gamma_{\rm op}$. Therefore, we expect that the excess photons can be substantially reduced to $ N_{\rm ex}(\Gamma^{-1}) \,\Gamma/\gamma_{\rm op}$ by appropriate filtering, which cancels the large factor $\gamma_{\rm op}^2/\lambda^2$ appearing in \eeqref{eq:Nex}. Indeed, we find below that the OMT at a second node provides exactly this filtering.

\section{Quantum Networks}
\label{sec:longDistance}

With the concept of the OMT laid out in the previous section we are now ready to consider a quantum network, where the OMTs in the various nodes serve to link the qubits to a common optical fiber, as depicted in \figref{fig:overview}. To obtain an effective description of the multiqubit dynamics, we proceed as before by linearizing the OM couplings and eliminating the coupled OMTs. We will then discuss how to realize a state-transfer within a two-node network.
The Hamiltonian for the complete network is a straight-forward generalization of \eeqref{eq:HsingleNode} and reads:
\begin{align}
H=\sum_i \left( H^i_{\rm node} +H_\textrm{cav-fib}^i +H^i_{\rm env}\right) + H_{\rm fib} \,,
\end{align}
where $H^i_{\rm node}$ is given by \eeqref{eq:Hnode} with appropriate node indices added and $H_{\rm env}^i$ summarize decoherence effects. We assume that the output of a given cavity is routed to the next, which is, e.g., realized by side-coupling all cavities to the right-moving field in the fiber as described by
\begin{align}
H_\textrm{cav-fib}^i &=  i \sqrt{2\kappa^i_\smf}  \left( c_i f_R^\dagger(z_i) - c_i^\dagger f_R(z_i) \right)\,.
\end{align}
Here, $f_R(z_i)$ is the right-moving field operator defined after \eeqref{eq:HcavfibSingleNode} and the nodes are located at positions $z_i<z_{i+1}$ along the fiber, with $c_i$ denoting the driven cavity mode at node $i$.

In a first step, we eliminate the modes of the optical fiber in a procedure similar to the one in \secref{sec:QLEs}. For each node, we obtain an input-output relation of the form \eqref{eq:cavInOut} and in addition,  we find a cascaded coupling where the output of a given cavity drives the subsequent one \cite{Gardiner1993,*Carmichael1993,QuantumNoise}:
\begin{align*}
\dot c_i(t) 
&\approx i [H_{\rm node}^i,c_i]  -\kappa_f c_i(t) - \sqrt{2\kappa_\smf} f_{{\rm in},i}(t)\,,\\
f_{{\rm in},i}(t)&\approx f_{{\rm out},i-1}(t-\tau_{i,i-1}) e^{i\om_L \tau_{i,i-1}}\,.
\end{align*}
Here, $H_{\rm om}^i$ in $H_{\rm node}^i$ is replaced by the analogue of \eeqref{eq:HomBar}, $\tau_{ij}=(z_i-z_j)/c$ is the propagation time between the nodes and $f_{{\rm in},1}(t)=f_{\rm in}(t)$ is a vacuum white-noise operator. To simplify notation, we assume quantities without index to be the same for all nodes and define rotated and retarded photonic operators, i.e., $c_i(t)\rightarrow e^{i\om_L z_i/c}c_i(t-z_i/c)$ and similarly for $f_{{\rm in},i}$, $f_{{\rm out},i}$, $\mE_i$. All other quantities are redefined as, e.g., $b_i(t)\rightarrow b_i(t-z_i/c)$ and we can then shift the time in the QLEs for node $i$ according to $t\rightarrow t+z_i/c$. Including local dissipative effects as above, the full OM QLEs for $\lambda\rightarrow 0$ read
\begin{align}
\label{eq:cascadedLangevin}
\dot b_i(t) 
&= i[H_{\rm node}^i,b_i] - \frac{\gamma_m}{2}b_i -\sqrt{\gamma_m}\xi_i(t)\,,\\
\dot c_i(t)
&=i [H_{\rm node}^i,c_i] - \kappa c_i -\sqrt{2\kappa_\smf} f_{{\rm in},i}(t) - \sqrt{2\kappa_0}f_{0,i}(t)\nonumber\,,
\end{align}
where the cascaded coupling is encoded in the input-output relation
\begin{align}
\label{eq:cascInOut}
f_{{\rm in},i+1}(t)&=f_{{\rm out},i}(t)=f_{{\rm in},i}(t)+\sqrt{2\kappa_\smf}\, c_{i}(t)\,.
\end{align}

Now, we can treat the classical laser drives by making the replacements $c_i\rightarrow c_i+\alpha_i$ and $b_i\rightarrow b_i + \beta_i$, and from the requirement of vanishing classical forces on the shifted operators we obtain the system
\begin{subequations}
\label{eq:cascClassResp}
\begin{align}
\dot\alpha_i&=-(i\Delta_c+\kappa)\alpha_i -ig_0 (\beta_i+\beta_i^*)\, \alpha_i + \mE_i^\eff\,,\\
\dot\beta_i &=-(i\om_r + \gamma_m/2) \beta_i -ig_0\abs{\alpha_i}^2\, .
\end{align}
\end{subequations}
Here,  
$\mE_i^\eff=\mE_i - 2\kappa_f\sum_{j<i}\alpha_j$
is the effective coherent drive for node $i$, which is the superposition of the local drive and the output of previous cavities. By taking this accumulation into account, it is in principle possible to control the classical fields in the cavities independently. However, by considering schemes with more than one cavity per node \cite{Safavi-Naeini2011,Stannigel2010}, one could also prevent the classical control signals from propagating down the fiber in the first place and thereby achieve independent tunability more easily. The steady state for each $\alpha_i$ is given by \eeqref{eq:adiabaticE} with the replacement $\mE\rightarrow\mE_{i}^\eff$ and similar statements on adiabaticity apply.
Provided that $\sabs{\alpha_i}^2\ll 1$, we may linearize the OM coupling as before and the resonators and cavities are then described by the linear system
\begin{align}
\label{eq:linQLEcasc}
\dot {\vecc{v}}^i(t) = &-M^i \vecc{v}^i(t) - \vecc{R}^i(t) - \sqrt{2\kappa_\smf}\,\vecc{I}^i(t) \,.
\end{align}
where $\vecc{v}^i=(b_i,c_i,b_i^\dagger,c_i^\dagger)^T$ and $M^i$ is defined as in \eeqref{eq:OMmatrix}. The vector $\vecc{I}^i(t)=(0,f_{{\rm in},i},0,f^\dagger_{{\rm in},i})^T$ contains the fiber-input of node $i$, while $\vecc{R}^i(t)=(\sqrt{\gamma_m}\xi_i,\sqrt{2\kappa_0}f_{0,i},\sqrt{\gamma_m}\xi^\dagger_i,\sqrt{2\kappa_0}f^\dagger_{0,i})^T$ summarizes local noise inputs.

\subsection{Ideal effective qubit network}
\label{sec:idealQubitNetwork}

We derive an effective description of the qubits on the basis of the assumption that their coupling to the resonators is slow compared to the OM dynamics (cf. \secref{sec:singleNode}). This allows us to eliminate the OM degrees of freedom and most importantly, we expect the qubits to inherit the cascaded nature of the coupling present between the cavities. We have seen before that a single OMT mediates a qubit-fiber coupling $\sim \sqrt{\Gamma}$, and we hence expect the effective cascaded coupling between qubits $i$ and $j$ due to emission and re-absorption of a photon to scale as $\sqrt{\Gamma_i \Gamma_j}$. 
The general elimination procedure is presented in \appref{sec:appGeneralElimination} and \ref{sec:appLongFiber} and for later convenience in deriving the state transfer protocol and discussing imperfections, we present the results in a master equation (ME) formulation for the reduced qubit density operator $\mu$.

To focus on the key points, we postpone the general situation to the next subsection and first discuss the  idealized case in which we (i) take the OM coupling in RWA ($\zeta=0$) and (ii) assume all additional decay channels and noise sources to be zero. The general effective ME \eqref{eq:effCascMEapp} then assumes a very simple form: 
\begin{align}
\label{eq:MEideal}
\dot \mu &= \mL_0\mu + \mL_{\rm ideal} \mu\,.
\end{align}
Here,  $\mL_0\mu=-i \sum_i \tilde\om_q^i[\sigma_z^i,\mu]/2$ is the renormalized free qubit Liouvillian with  $\tilde\om_q^i=\om_q+\Delta_{0,i}$, where the shifts $\Delta_{0,i}$ are defined below. The second term describes the ideal cascaded interaction \cite{QuantumNoise} and reads
\begin{align}
\label{eq:Lideal}
\mL_{\rm ideal} \mu&= -i H_\eff \mu + i \mu H_\eff^\dag + \mS \mu \mS^\dag\,,\\
\label{eq:HeffCasc}
H_\eff 
&= -\frac{i}{2} \mS^\dag \mS 
-\frac{i}{2} \sum_{i>j} \sqrt{\Gamma_i\Gamma_j}\left(\sigma^+_i\sigma^-_j - \sigma^-_i\sigma^+_j\right)\,,
\end{align}
where $\mS=\sum_i \sqrt{\Gamma_i} \sigma_i^-$ is the collective jump operator and the single-qubit decay rates $\Gamma_i$ are given below. 
The first, anti-hermitian term in the effective Hamiltonian $H_\eff$ ensures that $\mL_{\rm ideal}$ is of Lindblad form with a single jump operator $\mS$ and the second, hermitian term describes the coherent part of the fiber-induced dynamics. In writing the above equations, we have further absorbed phases $e^{-i\theta_i}\sigma_i^-\rightarrow\sigma_i^-$ into the qubit operators to simplify notation 
\footnote{The two-qubit terms are determined by the cascaded coupling $J_{ij}$ introduced in \eeqref{eq:cascadedCoeffs}.  In OM RWA and for $\gamma_m=\kappa_0=0$ it can be written as $J_{ij}=\sqrt{\Gamma_i\Gamma_j}\exp[i(\theta_i-\theta_j)]$ for $i>j$, where   $\theta_i=\phi_i+\sum_{n=1}^{i-1}2\phi_n$ with $\phi_n={\rm arg}\{i A_{21}^n(\om_q)\}$ and  $G_i$ taken to be real for simplicity. 
This structure allows to absorb $e^{-i\theta_i} \sigma_i^- \rightarrow \sigma_i^-$.
}.
Note that the two-qubit terms in the hermitian and non-hermitian parts of  $H_\eff$ have the same magnitude and a specific phase-relation, such that they actually interfere to produce the cascaded coupling. This is more evident when rewriting $H_\eff$ as
\begin{align}
H_\eff &= -i \sum_i\frac{\Gamma_i}{2}\,\sigma^+_i\sigma^-_i -i \sum_{i>j}\sqrt{\Gamma_i\Gamma_j}\, \sigma^+_i\sigma^-_j \,.
\end{align}
Apart from generating single-qubit decays, this Hamiltonian may transfer excitations from qubit $j$ to a subsequent qubit $i>j$, but not in the other direction. As expected, these cascaded interactions take  place on the scale $\sqrt{\Gamma_i\Gamma_j}$ and we will exploit them to perform a state transfer below.

The effective quantities entering the above ME are given by the dynamical properties of the underlying multinode OM system.  To determine the cascaded dynamics we use the general relations from  \appref{sec:appLongFiber},
\begin{align}
\label{eq:cascadedCoeffs}
\Gamma_i+2i\Delta_{0,i}=\frac{\lambda^2}{2} X_{ii}(\om_q)\,,\qquad
J_{ij}=\frac{\lambda^2}{4} X_{ij}(\om_q)\,,
\end{align}
where the cascaded coupling $J_{ij}$ determines the two-qubit terms in $H_\eff$. The correlation function $X_{ij}(\om)$ is generally defined as 
\begin{align}
X_{ij}(\om)&=\int_0^\infty\rmd\tau\,\smean{[b_i(\tau),b_j^\dagger(0)]}_{\rm free}\,e^{i\om\tau}\,,
\end{align}
where the subscript ``free" denotes the expectation value in the steady state of the OM dynamics in the absence of the qubits as described by Eqs.\,\eqref{eq:cascInOut} and \eqref{eq:linQLEcasc}.
It can be evaluated using the quantum regression theorem \cite{QuantumNoise} with the initial condition $\smean{[b_i(0),b_j^\dagger(0)]}_{\rm free}=\delta_{ij}$. Defining the OM response matrix of node $i$ as $A^i(\om_q)=(M^i-i\om\mathbb{1})^{-1}$, we find in general that $X_{ii}(\om)=A^i_{11}(\om)$ and also that $J_{ij}=0$ for $i<j$, which yields the uni-directionality of the coupling.
In contrast, for $i>j$ the correlation function describes the propagation of a signal from node $j$ to node $i$ including various filtering and absorption effects taking place at intermediate nodes (see \appref{sec:appLongFiber} for details). For the idealized case considered in this section, these effects only amount to a phase and we obtain $\abs{J_{ij}}=\sqrt{\Gamma_i\Gamma_j}$ as expected.

\subsection{Full master equation}
\label{sec:completeME}

\begin{figure}
\includegraphics[width=0.5\textwidth]{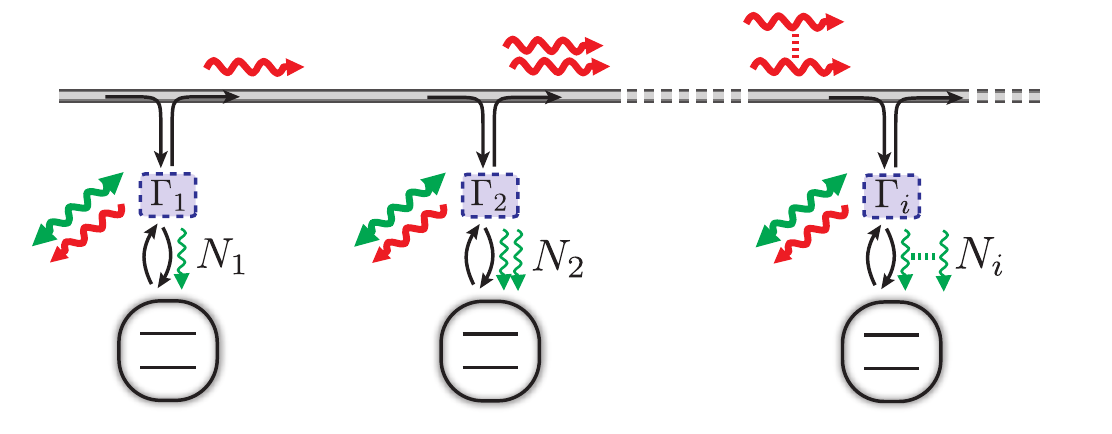}
\caption{(Color online) Accumulation of noise in a cascaded network where. Thick, wavy arrows denote noise photons (red, single-headed) and noise phonons (green, double-headed). Each node emits noise photons generated by non-RWA scattering events or by up-conversion of thermal phonons. At successive nodes, these noise photons may be down-converted and lead to decoherence of the qubits (thin wavy arrows).}
\label{fig:cascadedFigure}
\end{figure}

The ideal picture presented in the previous subsection has to be refined in order to discuss the impact of the various imperfections. As has already been discussed in Sec\,\ref{sec:singleNode}, the effects of the thermal noise and non-RWA corrections for a single node are two-fold: They lead to heating of the attached qubit as well as contamination of the OMT's output with noise photons. In a multiqubit network, these noise photons naturally affect successive nodes and when moving down the fiber we expect an accumulation of noise, as illustrated in \figref{fig:cascadedFigure}. 
The general ME for an $N$-node network capturing all these imperfections is derived in \appref{sec:appLongFiber} and reads:
\begin{align}
\label{eq:effCascME}
\dot\mu 
&= \frac{1}{2}\sum_{i}\Big\{ 
-i  \tilde\om_q^i [\sigma_z^i,\mu] + \Gamma_i \mD[\sigma^-_i ] \mu \Big\} \\
&\phantom{=}- \sum_{i>j}\left(J_{ij}\,\left[\sigma^+_i,\sigma^-_j\mu\right]+J_{ij}^*\,\left[\mu\sigma^+_j,\sigma_i^-\right]\right)\nonumber\\
&\phantom{=}+ \frac{1}{2}\sum_i\Gamma_i N_i \left( \left[\left[\sigma_i^+, \mu\right],\sigma_i^-\right] + \left[\sigma^+_i,\left[\mu,\sigma^-_i\right]\right]\right) \nonumber\\
&\phantom{=}+ \sum_{i\neq j}  D_{ij} \left[\left[\sigma_j^+,\mu\right],\sigma_i^-\right] \,, \nonumber
\end{align}
where $\mD[a]\mu=2a\mu a^\dag -a^\dag a \, \mu - \mu \, a^\dag a$ is a Lindblad term with jump operator $a$. 
The first two lines contain the fiber-induced decay and cascaded coupling of the qubits as discussed in the previous subsection, with the associated rates $\Gamma_i$ and $J_{ij}$ given in \eeqref{eq:cascadedCoeffs}. The renormalized qubit frequencies are given by $\ti\om_q^i=\om_q+\Delta_{i}$, with $\Delta_i=\Delta_{0,i}+\Delta_{{\rm th},i}$, where the second, thermal contribution $\Delta_{{\rm th},i}$ is negligible for our purposes (see the discussion in App\,\ref{sec:appLongFiber}). In contrast, the last two lines describe additional decoherence processes, characterized by effective bath occupation numbers $N_i$ and the rates of  correlated diffusion $D_{ij}$. These quantities are given by
\begin{align}
\Gamma_i N_i= \frac{\lambda^2}{4} Y_{ii}(\om_q)\,, \qquad
D_{ij}=\frac{\lambda^2}{4} Y_{ij}(\om_q)\,,
\end{align}
where $Y_{ij}(\om)$ is determined by the OM dynamics according to
\begin{align}
Y_{ij}(\om)&=\int_{-\infty}^\infty\rmd\tau\,\smean{b_i^\dag(\tau) b_j(0)}_{\rm free}\,e^{-i\om\tau}\,,
\end{align}
and the property $Y_{ij}(\om)=Y^*_{ji}(\om)$ ensures that the $N_i$ are real. Finally, we note that the explicit Lindblad form of \eeqref{eq:effCascME} is in general not very illuminating and also tedious to compute except for special cases as given below or in the previous subsection. 
Nevertheless, it is ensured by the property $\smean{b_i^\dag(\tau) b_j(0)}_{\rm free}=\smean{b_i^\dag(0) b_j(-\tau)}_{\rm free}$ of the OM correlation functions that the  qubit density operator $\mu$ remains positive semi-definite under the evolution described  by \eeqref{eq:effCascME} (see Ref.\,\cite{BreuerPetruccione2002}).

In the previous subsection, we assumed the idealized conditions of vanishing intrinsic decays $\kappa_0=\gamma_m=0$, and $\zeta=0$. In this limit, we obtain $N_i=0$, $D_{ij}=0$, and $\abs{J_{ij}}=\sqrt{\Gamma_i\Gamma_j}$, which allows one to rewrite the result \eeqref{eq:effCascME} in the  simple form of \eeqref{eq:MEideal}. We will now relax these restrictions to discuss the influence of the various imperfections and give the conditions under which their effects are small.

\subsubsection{Intrinsic cavity decays}

For non-vanishing intrinsic cavity decay $\kappa_0$ only a portion $\eta=\kappa_f/\kappa$ of the qubit excitation is actually emitted into the fiber, while the rest gets lost to other channels. For the simple case of two nodes and $\zeta=0$, the ME can be written exactly as the sum of a reduced cascaded interaction plus additional on-site decays: 
\begin{align}
\dot\mu &= \mL_0\mu + \eta \mL_{\rm ideal} \mu 
+ (1-\eta)\sum_{i=1}^{2}\frac{\Gamma_i}{2}\mD[\sigma_i^-]\mu\,,
\end{align}
with $\Gamma_i$ given in \eeqref{eq:decayRate}. For more than two nodes the ME is less simple, since every cavity provides an additional decay channel with an associated jump operator involving all previous qubits. Therefore, we only note that the cascaded couplings decay exponentially with the number of intermediate nodes, i.e., $\abs{J_{ij}}=\sqrt{\Gamma_i}\abs{t_{i-1}\ldots t_{j+1}}\sqrt{\Gamma_j}$ where $t_i(\om_q)=C_{22}^i(\om_q)$ are transfer amplitudes of the intermediate cavities with the matrices $C^i(\om)$ defined in \appref{sec:appLongFiber}. For $\kappa_0\ll\kappa_f$ and $\Delta_c=\om_r$ we find $1-\abs{t_i}\lesssim 2\kappa_0/\kappa_f$.

\subsubsection{Non-RWA corrections}

The most important effect of the non-RWA terms in the OM coupling is the appearance of additional contributions to the single- and multiqubit noise terms due to Stokes scattering, as has already been discussed for the single-qubit case in \secref{sec:singleNode} (see \eeqref{eq:N0}). Since explicit expressions for these contributions are lengthy and not very illuminating, we only mention that their effect scales as $\kappa^2/\om_r^2$, $\sabs{G}^2/\om_r^2$ and they are therefore small in the resolved-sideband limit.
Note, however, that the non-RWA terms in the various nodes are in principle phase-coherent, which can lead to non-trivial interference effects. Apart from the additional noise, one also obtains many rather quantitative corrections of the decay rates and cascaded couplings as compared to the RWA case, such as the fact that the relation $\abs{J_{ij}}\approx\sqrt{\Gamma_i\Gamma_j}$ becomes approximate. However, the general picture of emission and reabsorption of qubit excitations as conveyed by the ideal cascaded Liouvillian $\mL_{\rm ideal}$ remains valid.

\subsubsection{Accumulation of noise}

In \secref{sec:singleNodeImperfections} we have discussed the photonic noise present in the output of a single OMT. Clearly, in the current multinode setting these noise photons will propagate down the fiber and affect successive qubits (see \figref{fig:cascadedFigure}). 
Since the local noise sources (encapsulated in the $\vecc{R}_i$(t) in \eeqref{eq:linQLEcasc}) are uncorrelated, we expect the effective bath occupation numbers to scale with the number of preceding nodes, i.e., $N_i\propto i$.
To be more specific, we approximate them as
\begin{align}
\label{eq:noisePartitioning}
N_i \approx N_{0,i} + N_{c,i}\,.
\end{align}
Here, $N_{0,i}$ is the local contribution as given before in \eeqref{eq:N0} and the effect of previous nodes is summarized by the cascaded occupation number $N_{c,i}$. Neglecting a non-RWA pre-factor to $N_{c,i}$, it is defined as the spectrum of the cavity input evaluated at the qubit frequency, i.e., $N_{c,i}=\int_{-\infty}^\infty\rmd\tau\,\smean{f_{{\rm in},i}^\dagger(\tau)f_{{\rm in},i}(0)}_{\rm free}\,e^{-i\om_q \tau}$ and we have $N_{c,1}=0$ for the first node.

More explicitly, for the case of $N=2$ nodes, the cascaded noise is  simply determined by the output of the first cavity and it can be expressed as $N_{c,2}=2\kappa_f\,  \mC_{22}^{11}(\om_q)$,
with the steady-state correlation matrix $\mC^{11}(\om)$ defined in \appref{sec:appLongFiber}. Assuming that the first qubit is on resonance with one of its associated OM normal modes, we obtain for $\kappa,\abs{G}\ll\om_r=\Delta_c$:
\begin{align}
N_{c,2}\approx \frac{\kappa_f}{\kappa}\left(2 \frac{\gamma_m N_m}{\gamma_{\rm op}}+\frac{\kappa}{\gamma_{\rm op}}\frac{\sabs{G}^2}{\om_r^2}\right)\chi\,,
\end{align}
where $\chi=1$ for weak OM coupling ($\abs{G}\le\kappa/2$) and $\chi=\frac{1}{2}\abs{G}^2/(\abs{G}^2-\frac{3}{16}\kappa^2)\le 2$ for strong OM coupling ($\abs{G}>\kappa/2$).
When comparing this to the naive estimate for the excess noise given in \eeqref{eq:Nex}, we note that we have obtained the scaling $N_{c,2}\sim  N_{\rm ex}(\Gamma^{-1})\,\Gamma/\gamma_{\rm op} \sim N_{0,1}$. This means that the OMT at the second node provides exactly the filtering needed to suppress the noise originating from the first node (cf.  \secref{sec:singleNodeImperfections}). Therefore, we conclude that exchanging a single photon between two nodes of the quantum network is in principle possible.

For completeness, we briefly consider the general case of many nodes. To get a quantitative picture, it is again instructive to consider the simple case of $\zeta=0$ and $\kappa_0=0$, such that for $\gamma_m\rightarrow 0$ the only imperfection is mechanical diffusion at rate $\gamma_m N_m$. In this case, \eeqref{eq:noisePartitioning} holds exactly and we obtain
\begin{align}
N_{c,i}= \sum_{n<i} 2\kappa_f \sabs{A_{21}^n(\om_q)}^2\, \gamma_m^n N_m^n  \,,
\end{align}
where the appearance of the response matrix element $A_{21}^n(\om_q)$ reflects the fact that the thermal noise is first filtered by the OMT at node $n$ before it propagates down the fiber to affect node $i$. On resonance, the terms in the sum scale as $\gamma_m^n N_m^n/\gamma_{\rm op}^n$ and we thus obtain the expected accumulation of thermal noise. For identical nodes, this means $N_{i}\sim i\,\gamma_m N_m$ and a similar statement holds for the multiqubit diffusion terms, i.e.,
$
D_{ij}\sim {\rm min} (i,j) \,  \gamma_m N_m\, \Gamma/\gamma_{\rm op} 
$.
We finally note that within the present approximations, all of these results can be expressed compactly by rewriting the last two lines of \eeqref{eq:effCascME} as a sum over the thermal noise sources:
\begin{align*}
\frac{1}{2}\sum_{n} \gamma_m^n N_m^n \left(\left[\left[\Xi_n^\dagger,\mu\right],\Xi_n\right]+\left[\Xi_n^\dag,\left[\mu,\Xi_n\right]\right]\right)\,.
\end{align*}
This expression is explicitly in Lindblad form and the jump operator associated with the thermal noise originating from node $n$ involves all subsequent qubits according to $\Xi_n =\frac{\lambda}{2} \sum_{i\ge n} (\mT_{11}^{in}(\om_q))^*\, \sigma^-_i$.
Here,  the  matrices $\mT^{in}(\om)$ describe propagation from node $n$ to node $i$ (see \appref{sec:appLongFiber}) and we obtain $\frac{\lambda}{2}\mT_{11}^{in} \sim \sqrt{\Gamma_i/\gamma_{\smash{\rm op}}^n}$ if the qubits are resonant with the OM normal modes.

\begin{figure}
\includegraphics[width=0.5\textwidth]{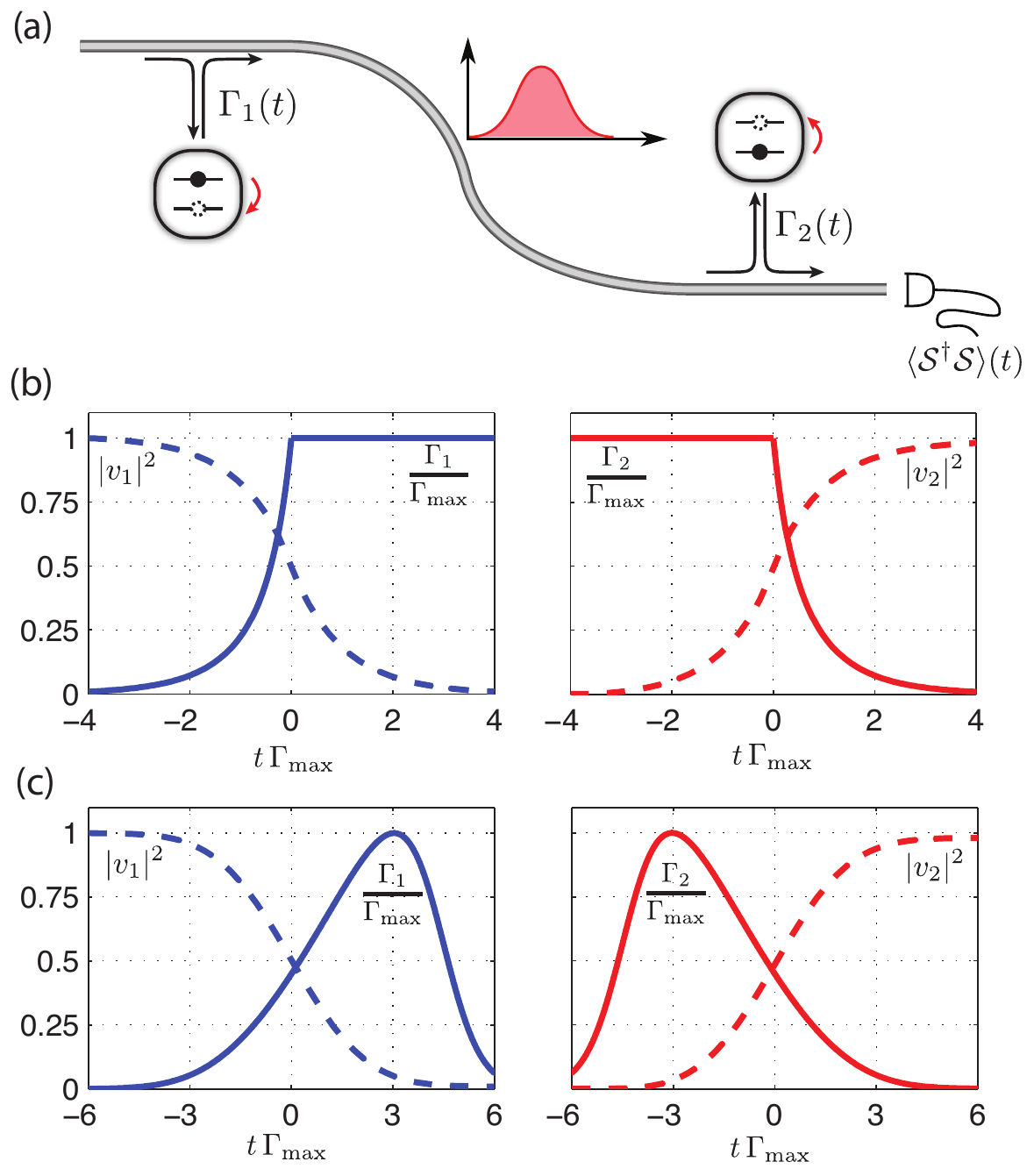}
\caption{(Color online) State transfer in a two-node cascaded quantum network. 
(a) Schematic illustration of time-symmetric photon wave-packet (see text). 
(b),(c) Exemplary pulses $\Gamma_i(t)$ given in \eeqref{eq:pulseShapes2} and \eeqref{eq:pulseShapes}, respectively, and resulting occupations $\abs{v_i}^2$ obtained numerically. In both cases, the parameters have been adjusted such that $\abs{v_1(t_f)}^2<10^{-2}$.}
\label{fig:stateTransfer}
\end{figure}

\subsection{State transfer protocol}
\label{sec:stateTransfer}

Up to now, the dynamics of the cascaded qubit network has been discussed on general grounds. We now turn to a specific application, namely the transfer of a quantum state from one qubit to the next. This has been the topic of our recent work \cite{Stannigel2010} and we provide here the details of the protocol. The main problem in transferring a quantum state $\ket{\psi_0}=\alpha\ket{0}+\beta\ket{1}$ between two nodes of a cascaded network according to
\begin{align}
\ket{\psi_0}_1\ket{0}_2 \rightarrow \ket{0}_1\ket{\psi_0}_2\,,
\end{align}
lies in the possibility that the photon emitted by the first node could pass 
the second node instead of being reabsorbed (see \figref{fig:stateTransfer}(a)). It was first realized by Cirac \etal \cite{Cirac1997} in the context of atomic cavity QED that such photon loss can be avoided by choosing appropriate time-dependent control pulses in the nodes, which leads to a deterministic state transfer protocol. In our setting, which is closely related, we can adiabatically tune the effective qubit-fiber couplings $\Gamma_i(t)$ as well as the bare qubit frequencies $\om_q^i(t)$.
We proceed by first deriving the ideal pulse-shapes needed for the state transfer and then comment on their realization.

\subsubsection{Pulse Shapes for state transfer}

We ignore all imperfections for the moment and base our derivation on the ideal cascaded ME as given in \eeqref{eq:MEideal} for $N=2$ qubits. Along the lines of Ref.\,\cite{Cirac1997} we require that the system remains in a pure state $\mu(t)=\pr{\psi(t)}$ during the entire evolution, which is equivalent to requiring the output of the photo-detector indicated in \figref{fig:stateTransfer}(a) to be exactly zero. It is clear from Eqs.\,\eqref{eq:Lideal},\eqref{eq:HeffCasc} that this is the case if we can enforce the so-called dark state condition $\mS(t)\ket{\psi(t)}=0$ for all times, where the time-dependence of the jump operator is attributed to the time-dependence of $\Gamma_i(t)$. Under this constraint the evolution of the system is completely characterized by the Schroedinger equation $\partial_t\ket{\psi}=-i(H_\eff + H_0) \ket{\psi}$, with $H_{\rm eff}$ given in \eeqref{eq:HeffCasc} and $H_0=\sum_i \tilde\om^i_q(t)\, \sigma_z^i / 2$ containing the (renormalized) qubit level splittings. We further re-introduce the phases $\theta_i$ that had been absorbed in \secref{sec:idealQubitNetwork} and  expand the state vector in terms of three time-dependent amplitudes $u$,$v_1$,$v_2$ as follows:
\begin{align} 
&\ket{\psi(t)}=\alpha u(t)\,e^{i\Phi_+(t)}\,\ket{00} \\ 
&+\beta\left[ v_1(t)\,e^{i\Phi_-(t)}\,\ket{10}+v_2(t)\,e^{i\Phi_-(t)+i\phi(t)}\,\ket{01} \right]\,.
\end{align}
Here, the phase factors have been introduced for later convenience according to $\Phi_\pm(t)=\frac{1}{2}\int_{t_0}^{t}\rmd s\, (\tilde \om_q^2(t) \pm \tilde \om_q^1(t))$ and $\phi(t)={\rm arg} \{J_{21}\}=\theta_2-\theta_1$. The dark-state condition
$\mS\ket{\psi}=0$ now reads $\sqrt{\Gamma_1}v_1 + \sqrt{\Gamma_2}v_2 = 0$
and the Schr\"odinger equation amounts to $\dot u=0$ and 
\begin{align}
\label{eq:schroedingerAmplitudes}
    \dot{v}_1 &= - \frac{\Gamma_1}{2}  v_1\,, \quad
    \dot{v}_2 = \left( -\frac{\Gamma_2}{2} - i\delta(t)\right) v_2 - \sqrt{\Gamma_1\Gamma_2} v_1\,,
\end{align}
where $\delta(t)=\tilde\om_q^2(t)-\tilde\om_q^1(t)+\dot\phi(t)$ is an effective detuning. We will first derive the pulses for $\delta(t)=0$, which is a generalized resonance condition taking the varying phase of the cascaded coupling into account.

Achieving perfect state transfer means that we find time-dependent $\Gamma_i(t)$ such that the solution of \eeqref{eq:schroedingerAmplitudes} satisfies the dark state condition as well as the boundary conditions
\begin{align}
\label{eq:boundaryCond}
v_1(t_i)=\abs{v_2(t_f)}=1\,,\qquad v_1(t_f)=v_2(t_i)=0\,,
\end{align}
where $t_i$ and $t_f$ are the initial and final times, respectively.
In reality, however, one will have to tolerate slight violations of these boundary conditions due to finite pulse lengths, etc.
To find suitable pulses we make use of the following time-symmetry argument \cite{Cirac1997,Korotkov2011}: If a photon is emitted by the first qubit, then, upon reversing the direction of time, we would see a perfect reabsorption. We can exploit this by ensuring that the emitted photon shape is invariant under time-reversal and use a time-reversed control pulse for the second qubit. As a result, the absorption process in the second node is a time-reversed copy of the emission in the first and may thus be -- at least in principle -- perfect. In the present formulation the role of the photon shape is played by $a(t)\equiv\sqrt{\Gamma_1}v_1$ and we can generate a class of pulses by requiring its time-derivative to be of the form $\dot a=f(t)a$ with some $f(t)$ satisfying $f(t)=-f(-t)$, provided we assume $t_f=-t_i$. Upon using \eeqref{eq:schroedingerAmplitudes} this yields a differential equation for the pulse shape:
\begin{align}
\label{eq:pulseShapeODE}
\dot\Gamma_1 = \Gamma_1^2+2f(t)\Gamma_1\,.
\end{align}
After choosing a suitable (i.e. bounded and positive) solution one may use it to calculate $v_1(t)$ and the quantities in the second node are then given by $v_2(t)=-v_1(-t)$ and $\Gamma_2(t)=\Gamma_1(-t)$, which solve the Schr\"odinger equation and satisfy the dark-state-condition. Note also, that related time-reversal arguments can be employed for reabsorbing almost arbitrary photon wave-packets, as has been discussed for formally similar atomic systems \cite{Gorshkov2007}.

We assume that there is a maximal achievable decay rate $\Gamma_\textrm{max}$ and give two example pulse-shapes. 
First, for $f(t)=-\Gamma_\textrm{max}\,\textrm{sign}(t)/2$ the wave-packet has the shape $a(t)\propto e^{-\Gamma_\textrm{max}\abs{t}/2}$ and a solution of \eeqref{eq:pulseShapeODE} is given by
\begin{align}\label{eq:pulseShapes2}
\Gamma_1(t<0)=\Gamma_\textrm{max}\frac{e^{\Gamma_\textrm{max}t}}{2-e^{\Gamma_\textrm{max}t}}\,,\qquad
\Gamma_1(t\ge0)=\Gamma_\textrm{max}\,.
\end{align}
The resulting dynamics is illustrated in \figref{fig:stateTransfer}(b), where the pulse-length has been chosen to be $T_p\equiv t_f-t_i=8/\Gamma_\textrm{max}$ such that $\abs{v_1(t_f)}^2<10^{-2}$. As a second example, we consider $f(t)=-ct$ for $c>0$ and obtain the solution
\begin{align}
\label{eq:pulseShapes}
  \Gamma_1(t)=\frac{\exp(- c t^2)}{\frac{1}{\Gamma_1(0)} - \frac{\sqrt{\pi}}{2\sqrt{c}} \textrm{Erf}\left(\sqrt{c}t\right)}
\end{align}
where the parameters $c$ and $\Gamma_1(0)$ have to be adjusted such that the boundary conditions \eqref{eq:boundaryCond} are met. Again, we display an example with $\abs{v_1(t_f)}^2<10^{-2}$ in \figref{fig:stateTransfer}(c). Note that the pulse-length $T_p=12/\Gamma_\textrm{max}$ is slightly longer than for the previous pulse, since the maximal decay is reached only shortly.

\subsubsection{State transfer between two nodes}

To discuss the implementation of the state transfer protocol in more detail, we focus on the pulse shape given in \eeqref{eq:pulseShapes}. We expect the other pulse (cf. \eeqref{eq:pulseShapes2}) to perform worse in terms of non-RWA noise due to the fact that the maximum decay is required over a long period. The following discussion is, however, valid in general.

The task is to tune the OM parameters in such a way that the resulting effective decays $\Gamma_i(t)$ in the two nodes follow the prescribed time-evolution $\tilde \Gamma_i(t)$, and we first consider varying the OM couplings $G_i(t)$ as depicted in \figref{fig:singleNode}(c).
The key step is then to solve the equation $\Gamma_i(G_i(t))=\tilde\Gamma_i(t)$ with $\kappa_0=\gamma_m=0$ for the desired temporal variation of the OM coupling $G_i(t)$, where we use the full relation \eeqref{eq:effCoeffs} to also capture non-RWA corrections. 
Note that the renormalization of $\Delta_c$ has to be taken into account
and we have further ignored the effective qubit shifts $\Delta_i$ as well as the varying phase of the cascaded coupling. This means that the resonance condition is in general not satisfied, i.e., we have $\delta(t)\neq0$. In order to correct this, we can adjust the bare qubit frequencies and in a first iteration set  $\om_q^i \rightarrow \om_q^i+\delta\om_q^i(t)$, with $\delta\om_q^{1}(t)=-\delta\om_q^{2}(t)=\delta(t)/2$. This shift is of order $\lambda^2/\gamma_{\rm op}$ and the new $\tilde\om_q^i$ approximately satisfy the resonance condition with corrections being of higher order in $\lambda$. In principle, shifting the bare qubit frequencies also modifies the $\Gamma_i(t)$ which therefore deviate from the prescribed $\tilde\Gamma_i(t)$. However, this deviation is negligible since the shifts $\delta\om_q^i$ are small compared to the width $\gamma_{\rm op}$ of the resonances in $\Gamma(\om_q,G)$ 
\footnote{To satisfy all of these constraints exactly, one may perform a minimization of the functional 
$\mC[\delta\om_q^i(t),G_i(t)]=\int_{t_i}^{t_f}\rmd t\,[(\delta(t))^2 + \sum_i(\Gamma_i(t)-\tilde\Gamma_i(t))^2]$.
}.
We have thus obtained the  pulse $(\delta\om^1_q(t),G_1(t),\delta\om^2_q(t),G_2(t))$ that realizes the desired control pulse $\tilde\Gamma_i(t)$ and satisfies the resonance condition. 

An example for a resulting pulse is displayed in \figref{fig:fidelitiesStateTransfer}(a) and we  note that  the adiabaticity conditions are well satisfied: Since $G(t)$ is tuned from 0 up to its maximum value in about one quarter of the pulse-length $T_p$, we have on average $\dot G/G_{\rm max}\sim 1/T_p\propto\Gamma_{\rm max}\ll \gamma_{\rm op}$, as is required for the adiabatic elimination to hold and is therefore valid by assumption. In addition, for translating the desired pulse $G(t)$ into applied laser power via \eeqref{eq:adiabaticE}, we need $\dot G/G_{\rm max}\ll\kappa$, which is implied by the previous relation. To assess the impact of imperfections we can now evaluate all quantities in the effective ME \eqref{eq:effCascME} for potentially finite $\gamma_m$, $\kappa_0$, $N_m$, as will be done below.
Finally, as an alternative to tuning the OM couplings $G_i$, we can also vary the cavity frequencies in order to regulate the effective decays, as depicted in \figref{fig:singleNode}(d) and a corresponding pulse $\Delta_{c,i}(t)$ is shown in \figref{fig:fidelitiesStateTransfer}(b).

\subsubsection{Discussion}

\begin{figure}
\includegraphics[width=0.5\textwidth]{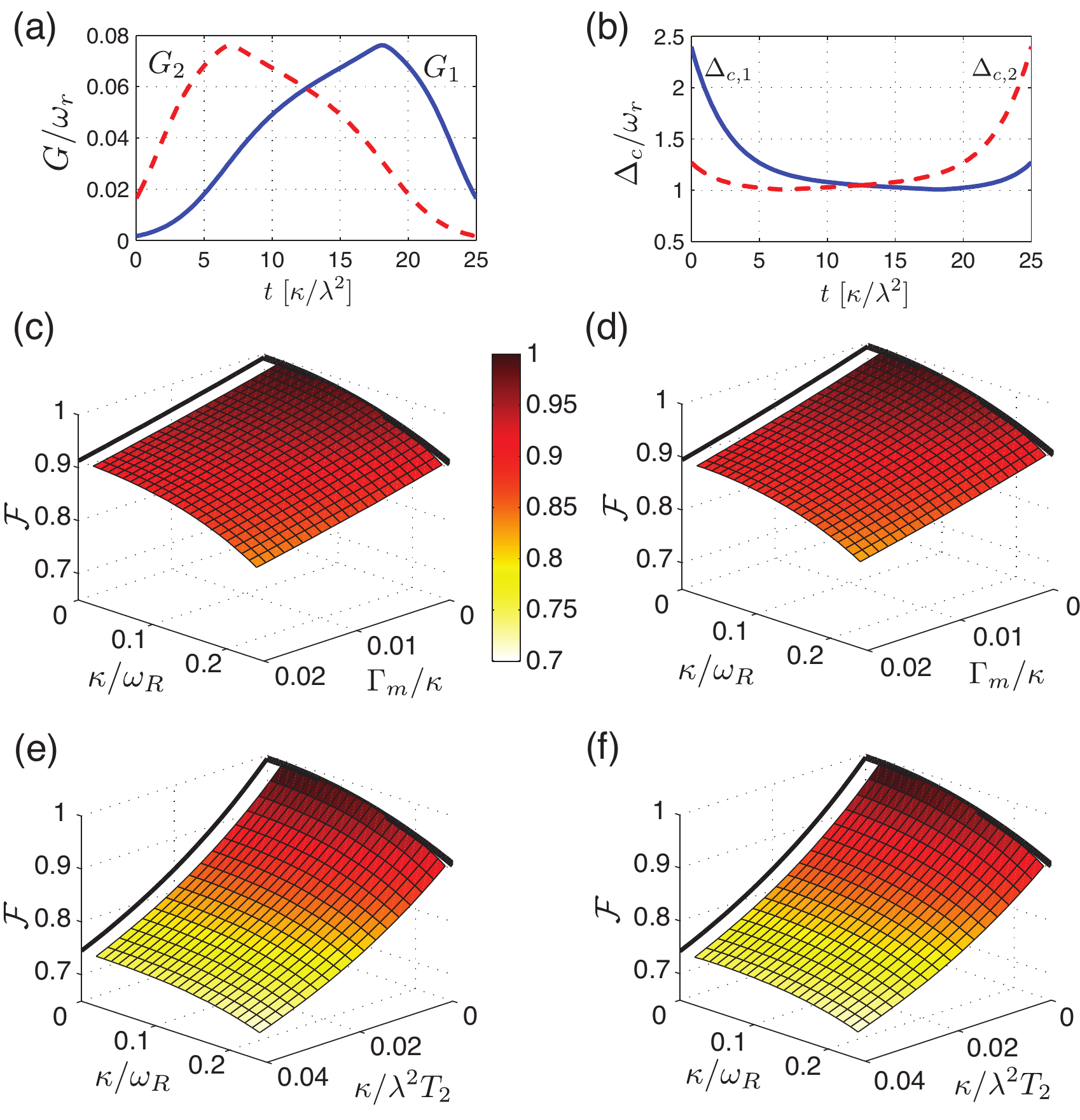}
\caption{(Color online) Performance of state transfer protocol using the pulse of \figref{fig:stateTransfer}(c). The left and right columns show realizations  via tuning of OM couplings and cavity frequencies, respectively.
(a),(b) Resulting control pulses for $\kappa=0.05\om_r$ (see also \figref{fig:singleNode}). 
(c),(d) Fidelities as functions of $\kappa$ and $\Gamma_m$, where $\kappa_0=1/T_2=0$. 
(e),(f) Fidelities as functions of $\kappa$ and $1/T_2$, with $\kappa_0=\Gamma_m=0$. For the left column we deduce the parameters of \eeqref{eq:stateTransferFid} to be $\mC_1\approx4$, $\mC_2\approx1.4$ and $\mC_3\approx 7.5$, and for the right column $\mC_2$ remains the same while the others are slightly worse, i.e.,  $\mC_1\approx 5$ and $\mC_3\approx8$.
}
\label{fig:fidelitiesStateTransfer}
\end{figure}

The quality of the state transfer can be estimated by averaging the fidelity
\begin{align}
\mF(\psi_0)=\bra{\tilde\psi_0} \tr_1\{\mu(t_f)\}\ket{\tilde\psi_0}
\end{align}
over all transmitted single-qubit states $\ket{\psi_0}=\alpha\ket{0}+\beta\ket{1}$.
Here, $\mu(t)$ is the solution of the ME \eqref{eq:effCascME} with initial condition $\mu(0)=\pr{\psi_0}\otimes\pr{0}$ and $\ket{\tilde\psi_0}=\alpha e^{i\Phi_+(t)} \ket{0}-\beta e^{i\Phi_-(t)+i\phi(t)} \ket{1}$ is the target state including the various phases that are accumulated  due to the varying qubit frequencies and phase of the cascaded coupling. We also take into account a decay of the qubit coherences according to $e^{-t/T_2}$ by adding a term $\frac{1}{4T_2}\sum_i\mD[\sigma_z^i]\mu$ to the ME, where $T_2$ is the qubit's intrinsic coherence time. The leading contributions to the fidelity are expected to scale as
\begin{align}
\label{eq:stateTransferFid}
\mF\approx1-\frac{2}{3}\frac{\kappa_0}{\kappa} -\mC_1 \frac{\gamma_m N_m}{\kappa} 
-\mC_2\frac{\kappa^2}{\om_r^2}-\mC_3\frac{\kappa}{\lambda^2 T_2}\,,
\end{align}
where the $\mC_i$ are numerical constants of order one and the four terms correspond to intrinsic cavity loss, thermal noise, Stokes scattering (we chose $G_{\rm max}\sim\kappa $ for the pulses), and intrinsic qubit dephasing. The constants $\mC_i$ depend on the pulse-shape and can be optimized depending on the experimental parameters. Figs.\,\ref{fig:fidelitiesStateTransfer}(c),(e) display the fidelities obtained in Ref.\,\cite{Stannigel2010} for varying $G_i(t)$, while Figs.\,\ref{fig:fidelitiesStateTransfer}(d),(f) show results for varying $\Delta_{c,i}(t)$. It can be seen that the two control schemes do not differ significantly, although we will use the former in our estimates below.

To conclude this section, we briefly discuss the results for an implementation with spin- and charge qubits, the details of which are described \secref{sec:realizations}. For spin qubits, we have $\lambda/2\pi$ on the order of $50\unit{kHz}$ if we choose $\om_r/2\pi\approx 5\unit{MHz}$ and we assume $\kappa/2\pi\approx1\unit{MHz}$ to be in the resolved side-band regime. With expected decoherence and decay rates $(\kappa_0,\Gamma_m)/2\pi=(50,10)\unit{kHz}$, as well as $T_2\approx10\unit{ms}$, we obtain state transfer fidelities of $\mF \approx 0.85$, where all imperfections contribute approximately equally. However, with recent improvements of $T_2$-times for  spin qubits \cite{Tyryshkin2011}, fidelities beyond $\mF\approx0.9$ seem feasible for our parameters.
In contrast, the fidelity for charge qubits is clearly dominated by intrinsic qubit dephasing. In this case, one can choose a larger resonator frequency $\om_r/2\pi\approx 50\unit{MHz}$ and  $\kappa/2\pi\sim\lambda/2\pi\sim 5\unit{MHz}$, whereas present-day technology allows for qubit dephasing times around $T_2\approx2\unit{$\mu$s}$. Again, we obtain $\mF\approx0.85$, but here, dephasing alone already accounts for an infidelity of $\sim 0.1$. In this case, strategies to overcome this problem would have a significant impact on the performance of the state transfer.
In particular, one could numerically optimize the pulse-shapes for gate-speed or depart from the adiabatic elimination and treat the full dynamics of the OMT, yielding faster gate-sequences. We conclude that state transfer operations with  both types of solid-state qubits are feasible with present-day technology.

\section{Optical on-chip networks}
\label{sec:onchip}

\begin{figure}
\includegraphics[width=0.5\textwidth]{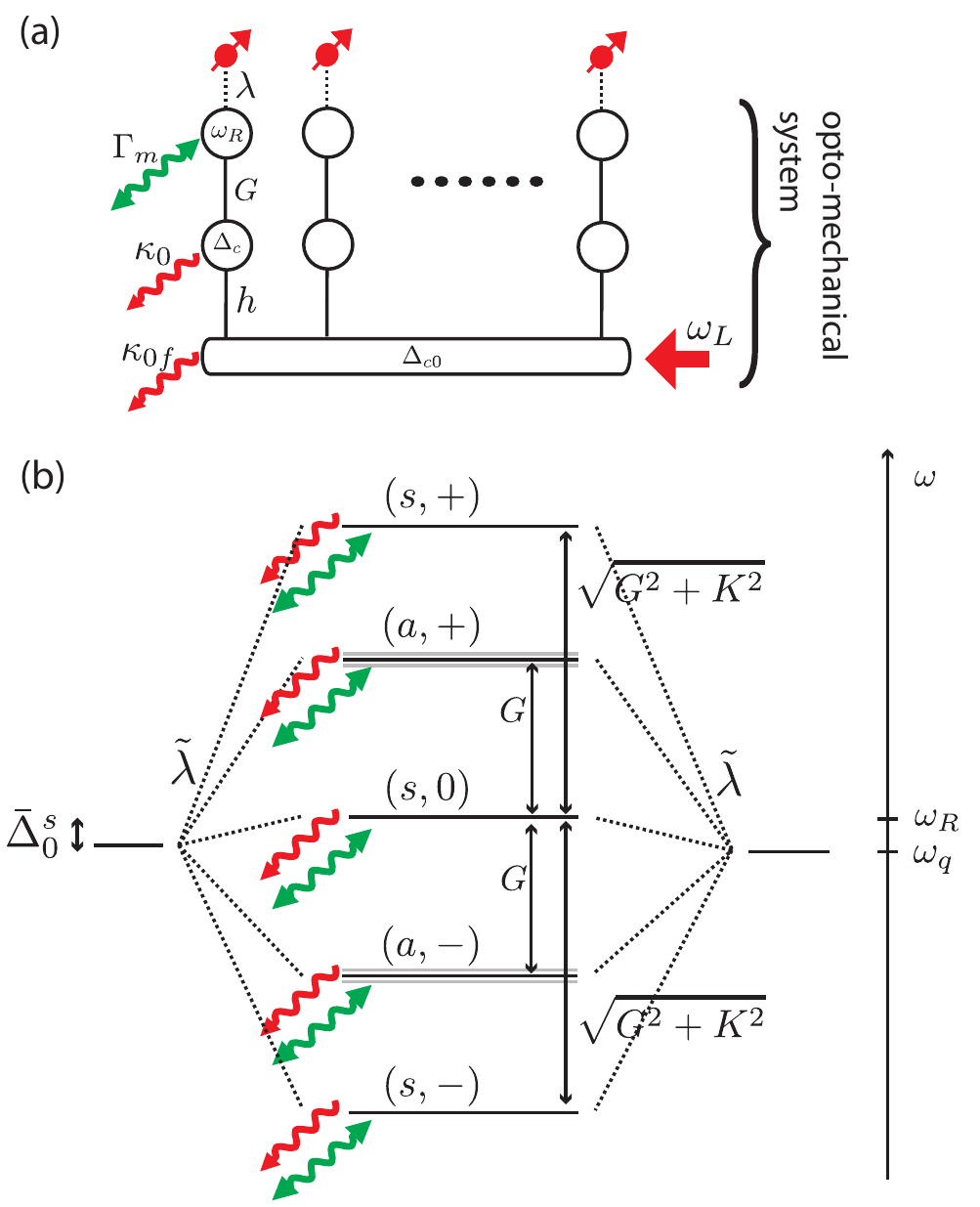}\\
\caption{(Color online) Phonon-photon bus realized via a fiber cavity. 
(a) Schematic illustration of the setup. Each circle represents an OM mode and the long box indicates the fiber cavity which is driven by a laser of frequency $\om_L$. Wavy arrows indicate cavity decay channels (red, single-headed) and mechanical diffusion (green, double-headed). 
(b) Level scheme for $N=2$ identical nodes  (black levels) corresponding to the results presented in Tab.\,\ref{tab:normalModes}. The two outer levels signify the two qubits to be coupled. For $N>2$ the modes labeled $(a,\pm)$ become $N-1$-fold degenerate (indicated in gray). The normal modes in RWA and for vanishing dissipation are given by 
$w_\pm^a=(\ti b_a\pm \ti c_a)/\sqrt{2}$, as well as 
$w_0^s=\frac{K}{\delta} \ti b_s - \frac{G}{\delta} c_0$
and
$w_\pm^s=\frac{1}{\sqrt{2}}(\frac{K}{\delta} \ti b_s \pm \ti c_s + \frac{G}{\delta} c_0)$, where $\delta=\sqrt{G^2+K^2}$.
}
\label{fig:onchipLevelScheme}
\end{figure}

So far we have considered networks where quantum information is transmitted through a continuum of modes supported by an optical fiber using pulse-shaping techniques. While this is a natural setting for long-distance quantum communication, one may also imagine situations where the different nodes of the quantum network are closely spaced and could instead be linked by optical resonators which exhibit a discrete set of modes. This scenario could be relevant for quantum communication and entanglement distribution within quantum processing architectures built on a chip, with  optical channels providing a fast and robust alternative to other communication schemes, as discussed in the introduction.
Therefore, we consider a setup as shown in \figref{fig:onchipLevelScheme}(a), where each qubit is linked via an OMT to a ``bus"-cavity of length $L$, which may, e.g., be implemented by a short optical fiber. 
Similar to the long-distance network considered above, the system is described by a Hamiltonian of the form
\begin{align}
H = \sum_i \left(H_{\rm node}^i +H_\textrm{cav-link}^i \right) + H_{\rm link} +H_{\rm env}\,,
\end{align}
where the structure of $H_{\rm node}^i$ is the same as in \eeqref{eq:Hnode}, with the exception that we do not apply laser drives to the nodes (see below).
The bus is modeled by an additional cavity with discrete modes separated by a free spectral range of $\delta\omega/2\pi  \gtrsim  1\unit{GHz}$, which corresponds to $L\lesssim 10\unit{cm}$. We take this spacing to be much larger than the other couplings and decays and therefore, we can restrict our discussion to a single fiber mode with destruction operator $d_0$ and frequency $\om_0$ close to the frequencies $\om^i_c$ of the nodal cavities. In this case, we obtain 
\begin{align}
\label{eq:Hlink}
H_{\rm link}&=\om_0 d_0^\dagger d_0 + i (\mE e^{-i\om_L t} d_0^\dagger - \mE^* e^{i\om_L t} d_0)\,,\\
H_\textrm{cav-link}^i &= h_{i} \, d_0^\dagger c_{i}  + h_{i}^* \, d_0 c^\dag_{i} \,,
\end{align}
where we have also included a laser of frequency $\om_L$ which coherently drives the fiber mode. 
Here, we have neglected additional degenerate modes in the nodes which do not couple to the fiber. They  will not be classically occupied  and  will thus not couple significantly to the mechanical resonance  (see \secref{sec:singleNodeLinearization}).

To proceed, we can follow the basic steps described in \secref{sec:longDistance}: First, we introduce the QLEs for the OM degrees of freedom to describe dissipative effects \cite{QuantumNoise}:
\begin{subequations}
\label{eq:onchipQLEs}
\begin{align}
\dot d_0 &= -i[d_0,H^\prime] - \kappa_{0f} d_0 -\sqrt{2\kappa_{0f}} f_{0}(t)\,,\\
\dot c_i &= -i[c_i,H^\prime] - \kappa_{0} c_i -\sqrt{2\kappa_0} f_{0,i}(t)\,,\\
\dot b_i &= -i[b_i,H^\prime] - \frac{\gamma_m}{2}b -\sqrt{\gamma_m}\xi_i(t)\,.
\end{align}
\end{subequations}
Here, $H^\prime=H-H_{\rm env}$ and $\kappa_0$, $\gamma_m/2$ are the decay rates of the cavities and resonators with corresponding noise operators  $f_{0,i}(t)$, $\xi_i(t)$ as introduced after Eqs.\,\eqref{eq:singleNodeQLE1},\eqref{eq:singleNodeQLE2}.
In addition, we have introduced a decay rate $\kappa_{0f}$ and vacuum noise operator $f_0(t)$ for the fiber mode.  Second, we determine  the classical steady state  by transforming all photonic operators to a frame rotating at the laser frequency and subsequently making the replacements $d_0\rightarrow d_0+\alpha_0$, $c_i\rightarrow c_i+\alpha_i$ and $b_i\rightarrow b_i + \beta_i$. Demanding the c-number contributions to the transformed QLEs to vanish then yields ($i>0$)
\begin{subequations}
\begin{align}
\dot\alpha_0&= -(i\Delta_{c0}+\kappa_{0f})\alpha_0 -i \sum_i h_i\alpha_i + \mE \,, \\
\dot\alpha_i&= -(i\Delta_{ci}+\kappa_{0})\alpha_i -i h_i^* \alpha_0  -i g_0 \alpha_i\, (\beta_i+\beta_i^*) \,,\\
\dot\beta_i &= -(i\om_r +\gamma_m/2) \beta_i -ig_0 \abs{\alpha_i}^2 \,,
\end{align}
\end{subequations}
where we have introduced the detunings of the cavities from the laser drive according to  $\Delta_{c0}=\om_0-\om_L$ and $\Delta_{ci}=\om^i_c-\om_L$. Below,  we are interested in the case of identical nodes, where the above system admits for a uniform steady-state solution with identical  $\alpha_i$ and $\beta_i$. For $\abs{\alpha_i}^2\gg 1$ we may now carry out the linearization of the QLEs \eqref{eq:onchipQLEs}, which amounts to the replacement 
\begin{align}
\label{eq:Hlin}
H^\prime\rightarrow  H_{\rm osc} +
\sum_i\left[ \frac{\om_q}{2}\sigma^i_z + \frac{\lambda}{2} \left(\sigma^-_i b_i^\dagger + \sigma^+_i b_i \right) \right]\,,
\end{align}
where $H_{\rm osc}$ is given by 
\begin{align}
\label{eq:Hosc}
H_{\rm osc}
&=\Delta_{c0} d_0^\dag d_0 + \sum_i\left( d_0^\dagger\,  h_{i} \,  c_{i} +\hc\right)\\
&+\sum_i \left[ \om_r b_i^\dagger b_i + \Delta_{ci} c_i^\dagger c_i 
+(G_i c_i^\dagger + G_i^* c_i)(b_i+b_i^\dagger)\right].\nonumber
\end{align}
Here, $G_i=g_0 \alpha_i$ is the enhanced OM coupling and 
we have redefined
$\Delta_{ci} - 2\sabs{G_i}^2/\om_r\rightarrow\Delta_{ci}$. In summary, we have thus obtained a system of qubits which is connected by a finite linear network of harmonic oscillators.

The linear oscillator network $H_{\rm osc}$ can in principle be used to resonantly propagate excitations from one qubit to another. However, this would require a certain degree of control over the various couplings to route an excitation to its target node and also has the drawback that the oscillators' decay channels have a strong effect.
For these reasons we will focus on an off-resonant scheme: The qubits chosen to take part in the gate operation are tuned close to one of the normal modes of the intermediate OM system described by $H_{\rm osc}$. If $\bar\Delta$ is the detuning from that mode they experience an effective interaction at rate $J\sim\lambda^2/\bar\Delta$, provided $\lambda\ll\sabs{\bar\Delta}$. On the other hand, a decay $\gamma$ of the chosen normal mode will lead to an induced qubit-decay $\Gamma\sim\lambda^2\gamma/\bar\Delta^2$, which  can be suppressed for larger detunings according to $\Gamma/J\sim \gamma/\bar \Delta$, as long as the intrinsic noise-processes of the qubit do not play a role (see below). To exploit this scaling, we need of course well-resolved OM normal modes.

\begin{table*}
\begin{tabular}{l||c|c|c|c|c|}
mode $(\nu,j)$ & $\om$ & $\gamma$ & $\tilde\lambda$ & $N_{\rm th}$ & $N_{\rm non-RWA}$ \\
\hline
\hline
(a,$\pm$) & $\om_r \pm G$ & $\displaystyle\frac{\gamma_m}{4} + \frac{\kappa_0}{2}$ &  
$\displaystyle\frac{1}{\sqrt{2}}\lambda$ & 
$\displaystyle\frac{\gamma_m N_m}{4\gamma^a_\pm}$ &
$\displaystyle\frac{G^2}{4(G\pm\om_r)^2}$\\
\hline
(s,0) &     $\displaystyle\om_r $ & 
$\displaystyle\frac{K^2}{2\delta^2}\gamma_m + \frac{G^2}{\delta^2}\kappa_{0f}$ & 
$\displaystyle\frac{K}{\delta}\lambda$ &
$\displaystyle\frac{K^2}{\delta^2} \frac{\gamma_m N_m}{2\gamma^s_0}$ & 
$\displaystyle \frac{\kappa_0 K^2}{4\kappa_{0f}\om_r^2}+\mO\left(\frac{K}{\om_r}\right)^4$\\
\hline
(s,$\pm$) & $\displaystyle\om_r \pm \delta$ & 
$\displaystyle\frac{G^2}{4\delta^2}\gamma_m +\frac{\kappa_0}{2} + \frac{K^2}{2\delta^2}\kappa_{0f} $ & 
$\displaystyle\frac{G}{\sqrt{2}\delta}\lambda$ &
$\displaystyle\frac{G^2}{2\delta^2} \frac{\gamma_m N_m}{2\gamma^s_\pm}$& 
$\displaystyle\frac{g^4 \kappa_0}{4 (\kappa_{0f} + \kappa_0 + g^2 \kappa_0 )} \frac{K^2}{\om_r^2} + \mO\left(\frac{K}{\om_r}\right)^3$ \\
\end{tabular}
\caption{Characterization of OM normal modes ($\delta=\sqrt{G^2+K^2}$): 
The different columns display eigenfrequencies $\om$ and decay rates $\gamma$ as obtained by diagonalizing $\tilde M^\nu$, as well as quantities entering the effective ME \eqref{eq:twoNodeMEonchipAppr} in the independent-mode approximation, i.e., effective couplings $\tilde\lambda$, thermal occupation numbers $N_{\rm th}$ and non-RWA occupation numbers $N_{\rm non-RWA}$. The first four columns were obtained in OM RWA and we have generally made use of $\kappa_0,\kappa_{0f},\gamma_m\ll G,K$ and assumed $\Delta_c=\Delta_{c0}=\om_r$ to simplify the expressions. The last column displays results to first order in the non-RWA terms (see text) and we show only leading terms in $G^2/\om_r^2,K^2/\om_r^2$ for $\gamma_m\rightarrow 0$ and $g\equiv G/K\lesssim 1$.
}
\label{tab:normalModes}
\end{table*}

\subsection{Two node network}

We illustrate the resulting interactions for $N=2$ nodes and postpone matters regarding the scalability to \secref{sec:largerNetworks}.  To begin with, we analyze the mode-structure of the OM system described by $H_{\rm osc}$, which will provide the basis for understanding the induced qubit dynamics. Assuming identical nodes, we introduce symmetric and anti-symmetric combinations of all system and noise operators (i.e. $\tilde c_{s/a}=(c_1\pm c_2)/\sqrt{2}$, etc.). 
The second term in \eeqref{eq:Hosc} then simply reads $K d_0^\dagger \tilde c_s+\hc$ with $K=\sqrt{2}h$ and the QLEs for the oscillator network (\eeqref{eq:onchipQLEs} with $H^\prime\rightarrow H_{\rm osc}$) decouple into a symmetric and an anti-symmetric set:
\begin{align}
\label{eq:QLEtwoNode}
\dot{\ti{\vecc{v}}}^\nu(t) = -\ti{M}^\nu {\ti{\vecc{v}}}^\nu(t) - {\ti{\vecc{R}}}^\nu(t) \quad \nu=s,a\, .
\end{align}
Here, the degrees of freedom have been grouped into $\ti{\vecc{v}}^s=(\ti b_s,\ti c_s,d_0,\ti b_s^\dag,\ti c_s^\dag,d_0^\dag)^T$ and $\ti{\vecc{v}}^a=(\ti b_a,\ti c_a,\ti b_a^\dag,\ti c_a^\dag)^T$, and the noise vectors $\ti{\vecc{R}}^\nu(t)$ are mutually uncorrelated. Explicit expressions for the $\ti{M}^\nu$ and $\ti{\vecc{R}}^\nu(t)$ can be found in \appref{sec:appShortFiber} and we take $G,K>0$ from now on since their phases can be absorbed into the cavity operators. The anti-symmetric set behaves like a normal OM system as discussed in \secref{sec:singleNode} and for the relevant strong coupling case $G\gg \kappa_0,\gamma_m$,  mechanical and optical modes hybridize to produce normal modes at $\om_\pm^a\approx\om_r\pm G$. Here and in what follows we assume $\Delta_c=\Delta_{c0}=\om_r$ and the results in the present subsection are obtained for OM RWA ($\zeta=0$). In contrast, the symmetric subsystem is an augmented OM system including also the fiber cavity with normal modes at $\om_0^s=\om_r$ and $\om_\pm^s\approx \om_r\pm\delta$, where $\delta^2=G^2+K^2$. Therefore, symmetric and anti-symmetric normal modes are staggered as shown in \figref{fig:onchipLevelScheme}(b) and for $G,K\gg\kappa_0,\kappa_{0f},\gamma_m$ the modes are well resolved, as can be seen from the expressions for their widths $\gamma$ listed in Tab.\,\ref{tab:normalModes}.

We proceed to eliminate the OM degrees of freedom on the basis of the assumption that they are detuned from the qubits by much more than $\lambda$. In terms of the new OM modes, the qubit-resonator interaction in \eeqref{eq:Hlin} reads
\begin{align}
H_{\rm int}=\frac{\lambda}{2}\sum_{\nu=s,a}\left( \ti\sigma_\nu^+ \ti b_\nu + \hc \right)\,,
\end{align}
and since the $\ti b_\nu$ are mutually uncorrelated they may be eliminated independently, as is done in \appref{sec:appShortFiber}. 
However, for well-resolved normal modes it is reasonable to further assume that also correlations between modes of the same set are small. To be specific, we introduce normal modes $\vecc{w}^\nu$ via $\vecc{v}^\nu=U^\nu \vecc{w}^\nu$, such that the transformations $U^\nu$ diagonalize  $\ti{M}^\nu$. This yields
\begin{align}
\label{eq:QLEnormalMode}
\dot w^\nu_j &= -( i \om_j^\nu+\gamma_j^\nu)w_j^\nu - \sum_k (U^\nu)^{-1}_{jk} \, \ti R_k^\nu \, ,
\end{align}
where $ i\om_j^\nu+\gamma_j^\nu$ are the eigenvalues of $\tilde M^\nu$. In addition, the qubit-resonator interaction becomes
\begin{align}
H_{\rm int}=\frac{1}{2} {\sideset{}{'}\sum_{\nu,j} } \left(\tilde\lambda_j^\nu \, \tilde\sigma_\nu^+ w^\nu_j +\hc \right)\,,\quad
\tilde \lambda_j^\nu = \lambda (U^\nu)_{1j}\, ,
\end{align}
where we have introduced the effective couplings $\lt_j^\nu$ and the primed sum is restricted to normal modes with frequencies $\om_j^\nu>0$. In the present RWA this is exact, but it presents an approximation otherwise.
Now, correlations between the normal modes $w_j^\nu$ of a given set $\nu$ arise due to the noise terms in \eeqref{eq:QLEnormalMode}, which scale as $\sqrt{\gamma_m}$, $\sqrt{2\kappa_{0f}}$, $\sqrt{2\kappa_0}$. Since these quantities have to be small anyway, we neglect these correlations and the normal modes $w_j^\nu$ then evolve independently. As a result, the general elimination procedure as outlined in \appref{sec:appGeneralElimination} becomes very simple and calculating the necessary correlation functions \eqref{eq:corrS}-\eqref{eq:corrY} for each $w_j^\nu$  yields a ME that is just a sum over the five normal modes listed in Tab.\,\ref{tab:normalModes}:
\begin{align}
\label{eq:twoNodeMEonchipAppr}
\dot\mu=
&-i \left[J(\sigma_+^1\sigma_-^2 + \sigma_-^1\sigma_+^2),\mu \right]
\\
+& \frac{1}{2}{\sideset{}{'}\sum_{\nu,j} } \Big\{
\Gamma^\nu_j (N^\nu_j+1)\mD[\ti\sigma^-_\nu]\mu 
+\Gamma^\nu_j N_j^\nu\mD[\ti\sigma^+_\nu]\mu
\Big\} \,, \nonumber
\end{align}
Here, $J=\sum_{\nu,j}^\prime J_j^\nu$ is the desired effective coupling and we have dropped terms renormalizing the qubit frequencies for simplicity, since they are the same for both qubits. The couplings and decay rates are directly related to the normal modes as described in Tab.\,\ref{tab:normalModes} and owing to our independent-mode approximation they have the form of simple Lorentzians ($\bar\Delta_j^\nu=\om_q-\om_j^\nu$):
\begin{subequations}
\label{eq:apprOnchipRates}
\begin{align}
J_j^\nu &= \frac{(\tilde\lambda_j^\nu)^2}{8} \, \frac{\bar\Delta_\nu^j (-1)^{\delta_{\nu a}}}{(\gamma_j^\nu)^2+(\bar\Delta_j^\nu)^2} \,, \\
\Gamma_j^\nu &= \frac{(\tilde\lambda_j^\nu)^2}{2} \, \frac{\gamma_j^\nu}{(\gamma_j^\nu)^2+(\bar\Delta_j^\nu)^2} \,.
\end{align}
\end{subequations}
Hence, we have obtained the expected scaling $\Gamma_j^\nu/J_j^\nu\sim\gamma_j^\nu/\bar\Delta_j^\nu$ when close to a resonance.
The alternating sign of the $J_j^\nu$ is rooted in the fact that eliminating a mode of symmetry $\nu$ yields a shift of the operator $\ti\sigma^+_\nu \ti\sigma^-_\nu\propto (-1)^{\delta_{\nu a}}(\sigma^+_1\sigma^-_2 + \sigma^-_1\sigma^+_2)$. Note that when tuning the qubits somewhere in between two modes, these have opposite symmetry (cf. \figref{fig:onchipLevelScheme}(b)) and their contributions to $J$  add up constructively, since the relevant detunings $\bar\Delta_j^\nu$ also have opposite signs.

Finally, the influence of the thermal noise on the qubits is characterized by the effective occupation numbers $N_j^\nu\sim\gamma_m N_m/\gamma_j^\nu$ (see Tab.\,\ref{tab:normalModes} for details). This means that the thermal noise is suppressed by an OM cooling effect, provided that we have $\gamma_m N_m \lesssim \kappa_0,\kappa_{0f}$.

\subsection{Non-RWA effects}
\label{sec:onchipNonRWA}

The effect of the non-RWA terms is to admix a counter-rotating portion to each OM normal mode, giving rise to new noise terms as well as quantitative corrections. In the case of the full OM coupling the eigenmode transformation $U^\nu$ is not block diagonal anymore and the eigenmodes in \eeqref{eq:QLEnormalMode} are driven by counter-rotating noise operators. To estimate their effect we can introduce new noise operators with effective non-zero occupation numbers by means of a Bogoliubov transformation and then proceed as above and treat the normal modes independently. For illustration, the eigenvectors of the anti-symmetric system $\tilde M^a$ are, to lowest order in the non-RWA parameter $\zeta$ and for vanishing dissipation,
\begin{align}
w_\pm^a
&\approx \frac{1}{\sqrt{2}}\left( \ti b_a \pm \ti c_a   \right) - \frac{\zeta G}{2\sqrt{2}(G\pm\om_r)}\left( \ti b_a^\dagger \pm \ti c_a^\dagger  \right) \,.
\end{align}
Calculating, also to first order in $\zeta$, the matrix $(U^a)^{-1}$ from these eigenvectors then allows to estimate the effective occupation number due to Stokes scattering:
\begin{align}
N_{\pm,{\rm non-RWA}}^a\approx
\frac{G^2}{4(G\pm\om_r)^2}\,,
\end{align}
where we have neglected non-RWA corrections to the thermal noise.
For the symmetric modes the expressions become quite lengthy, so we display in Tab.\,\ref{tab:normalModes} only the leading order terms in $G^2/\om_r^2$, $K^2/\om_r^2$.

\subsection{Discussion: $\sqrt{\textrm{SWAP}}$ gate}
\label{sec:onchipFidelity}

The effective two-qubit interaction obtained in \eeqref{eq:twoNodeMEonchipAppr} can be used to swap the states of the qubits or to perform the entangling $\sqrt{\textrm{SWAP}}$ operation, which amounts to exposing the qubits to this interaction for a time $t_g=\pi/4J$. 
To asses the optimal operation point for such operations we consider the creation of a maximally entangled state $\ket{\psi_0}=(\ket{01} - i \ket{10})/\sqrt{2}$ from the initial product state $\ket{01}$ by applying the $\sqrt{\textrm{SWAP}}$ gate.
For this purpose, we also account for a finite qubit coherence time $T_2$ by adding the term $\frac{1}{4T_2}\sum_i\mD[\sigma^i_z]$ to the ME \eqref{eq:twoNodeMEonchipAppr}, which is then solved to first order in the Lindblad terms. The resulting approximation of the fidelity $\mF=\bra{\psi_0}\mu(t_g)\ket{\psi_0}$ for this operation can be written as
\begin{align}
\label{eq:bellStateFid}
\mF \approx 1-\frac{\pi}{8} {\sideset{}{'}\sum_{\nu,j} }\frac{\Gamma_j^\nu}{\abs{J}}(1+2N_j^\nu) - \frac{\pi}{8}\frac{1}{\abs{J} T_2} \,,
\end{align}
which is a function of the qubit frequency (cf. \eeqref{eq:apprOnchipRates}).
Generally speaking, for $K\gtrsim G$ we expect higher fidelities if the qubits are tuned in between the central resonances (s,0), (a,$\pm$), as can be argued from Tab.\,\ref{tab:normalModes}: 
Compared to the outer resonances (s,$\pm$), they have larger weights $\ti\lambda^2$ and thus larger $J$. Further, their width $\gamma$ is slightly smaller for the relevant case   $\gamma_m\ll\kappa_0\approx\kappa_{0f}$.

In the limit where the induced decay $\Gamma$ dominates over intrinsic dephasing, i.e., $\Gamma\gg1/T_2$, we see from the second term in \eeqref{eq:bellStateFid}  that there is a point between each pair of neighboring resonances where $J/\Gamma$ is optimal. For $G=K$ such points can, e.g., be found half-way between resonances $(s,0)$ and $(a,\pm)$ at  $\om_q\approx\om_r\pm G/2$. By taking only the two closest resonances into account (which slightly over-estimates the fidelity) we obtain for $\kappa_0=\kappa_{0f}\ll G$
\begin{align}
\label{eq:fidOpt1}
\mF_{\rm opt}\approx 1 - \frac{\pi}{2} \frac{\kappa_0}{G}\left(1+  \frac{\gamma_m N_m }{\kappa_0} + \frac{G^2}{2\om_r^2}\right) - \frac{\pi}{2}\frac{G}{\lambda^2 T_2}\, .
\end{align}
In this limit, the infidelity is thus dominated by cavity-induced qubit decay (first term in parentheses), provided the thermal noise of the resonator is suppressed due to  $\gamma_m N_m\lesssim\kappa_0$. The last term in parentheses represents non-RWA corrections and is small for moderate couplings, as required by stability constraints (cf. \secref{sec:singleNode}).

In the opposite limit, where intrinsic qubit dephasing dominates ($1/T_2\gg\Gamma$), we have to tune very close to a single resonance $(\nu,j)$ to obtain large $J$ and hence a fast gate. There is, however, a trade-off between minimizing the dephasing (small $\bar\Delta$) and the optomechanically induced decay $\Gamma$ (large $\bar\Delta$). The optimal detuning in this case is given by $\bar\Delta_{\rm opt}\approx[(\lt_j^\nu)^2 \gamma_j^\nu (1+2 N_j^\nu)T_2/2]^{1/2}$ which yields a fidelity of
\begin{align}
\label{eq:fidOpt2}
\mF_{\rm opt}\approx 1-\pi\sqrt{\frac{2 \gamma_j^\nu(1+2 N_j^\nu)}{(\ti\lambda^\nu_j)^2T_2}}\,.
\end{align}
There is, however, also the limitation $\abs{\bar\Delta_{\rm opt}}\gg\lambda$ imposed by the elimination procedure.

\begin{figure}
\includegraphics[width=0.5\textwidth]{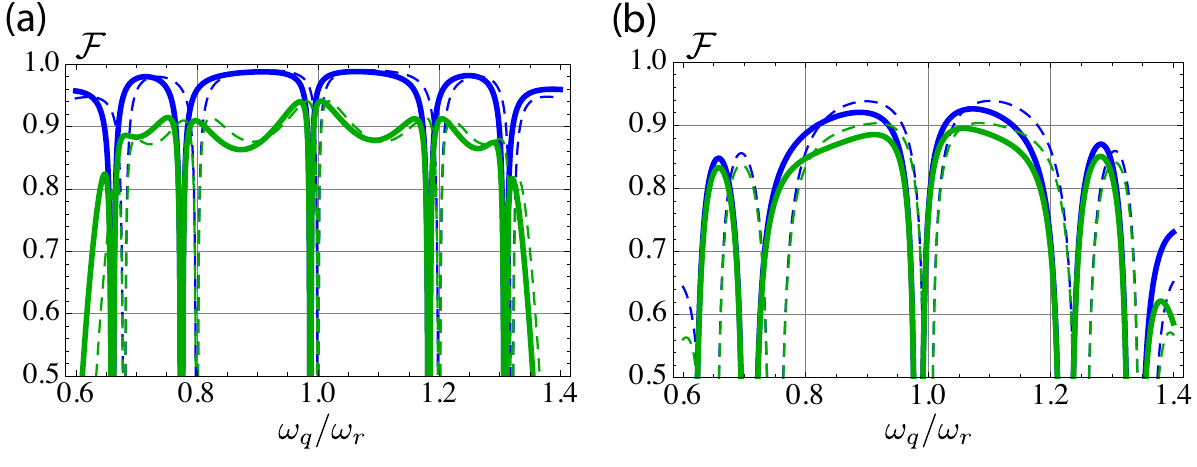}
\caption{(Color online) Fidelities for generating a maximally entangled state between $N=2$ nodes (see text). Solid curves: results of the full elimination presented in \appref{sec:appShortFiber}, dashed curves: approximate expressions according to \eeqref{eq:bellStateFid}. 
(a) Parameters suitable for charge qubit: $\om_r/2\pi=50\unit{MHz}$, $\lambda/2\pi=3.5\unit{MHz}$, $G=0.2\om_r$, $K=0.25\om_r$, and decoherence rates $(\kappa_0,\kappa_{0f},\Gamma_m)/2\pi=(50,25,10)\unit{kHz}$. The blue (dark gray) curves correspond to vanishing dephasing $1/T_2=0$ and the green (light gray) curves to $T_2=2\unit{$\mu$s}$.
(b) Parameters for spin qubit: $\om_r/2\pi=7.5\unit{MHz}$, $\lambda/2\pi=40\unit{kHz}$, $G=K=0.25\om_r$ and decoherence rates as in (a). Blue (dark gray) curves: $1/T_2=0$, green (light gray) curves: $T_2=10\unit{ms}$.
}
\label{fig:fidelitiesOnchip}
\end{figure}

Figure\,\ref{fig:fidelitiesOnchip} shows the fidelities for an implementation with spin and charge qubits as discussed in more detail in \secref{sec:realizations}. Apart from the  expression \eqref{eq:bellStateFid} based on the independent-mode approximation (dashed) we also display the result of the full elimination, \eeqref{eq:onchipFidelityFull}, showing reasonable agreement.
For an ideal superconducting qubit, the fidelities are rather good and with $T_2\approx 2\mu$s we still obtain $\mF\gtrsim0.85$ for the parameters of \figref{fig:fidelitiesOnchip}(a), which shows that the performance is limited by intrinsic dephasing.
On the other hand, for the spin qubit a resonator of lower frequency is better suited (yielding higher zero-point motion), which also puts limits on the OM coupling $G$. As a result, the leading term in \eeqref{eq:fidOpt1} proportional to $\kappa_0/G$ is larger than for the charge qubit, meaning that the intrinsic cavity decays become a more severe limitation in this case. For an ideal spin qubit we obtain fidelities of $\sim 0.9$ for parameters as in \figref{fig:fidelitiesOnchip}(b). Intrinsic dephasing then has a less severe impact than for the charge qubit  and for $T_2\approx10\unit{ms}$ we still find $\mF\approx0.85-0.9$. Strictly speaking, these estimates are only valid for small infidelities, but they are nevertheless useful for identifying interesting regimes of operation and we recognize the two limiting cases discussed above. As in the case of the long-distance state transfer, one might -- especially in the case of the charge qubit -- profit from leaving the adiabatic regime in order to shorten the gate-sequence and to reduce the impact of intrinsic qubit dephasing.

\subsection{Outlook: Scalability}
\label{sec:largerNetworks}

When considering an $N$-node network the most straight-forward way to perform a gate operation between two given qubits is to detune the nodal cavities of all other nodes sufficiently far, such that the network reduces to the two-node problem, which leads us to expect similar fidelities. Since this requires only moderate temporal control of the cavity frequencies $\om_c^i$, this could be achieved by applying, e.g., strain or heat to those cavities not meant to participate (see, e.g., Ref.\,\cite{Armani2004} for tuning of the toroidal cavities discussed in \secref{sec:realizations}). Note that if the fiber has to be extended to accommodate the new nodes, the cavity-fiber couplings scale as $h\propto1/\sqrt{L}$, which has to be taken into account.

We now consider $N$-qubit interactions and the basis for the discussion is again the normal-mode structure of the underlying $N$-node OM network. For identical nodes, only the ``center-of-mass" cavity mode $\ti c_1=\sum_i c_i/\sqrt{N}$ couples to the fiber while all other modes decouple. As a result, the level scheme in \figref{fig:onchipLevelScheme}(b) remains valid with the modification that the modes $(a,\pm)$ become $(N-1)$-fold degenerate (indicated in gray) and we assume that the fiber is extended with every node added, meaning  $h\propto1/\sqrt{N}$ and hence $K=\sqrt{N}h={\rm const}$. As could be expected, the calculation presented in \appref{sec:appShortFiber} shows that the induced coherent qubit dynamics is given by $J_N S^+S^-$, where $S^-=\sum_i\sigma^-_i$ and in general $J_N\propto 1/N$. For illustration, we note that if the qubits are tuned close to the symmetric mode $(s,0)$ at $\om_r$, we can neglect the anti-symmetric ones provided $G\gtrsim N \kappa_0$ and write the effective $N$-qubit ME \eqref{eq:MENnodes} as 
\begin{align}
\dot\mu \approx 
&-i\left[J_N S^+S^-,\mu\right] \\ 
&+\frac{\Gamma_{\rm coll}}{2}(N_{\rm coll}+1)\mD[S^-]\mu+\frac{\Gamma_{\rm coll}}{2} N_{\rm coll}\mD[S^+]\mu \,,\nonumber
\end{align}
where the coefficients are related to the two-node result via $J_N \sim 2J/N$ and  $\Gamma_{\rm coll}\sim 2\Gamma_0^s/N$, and $N_{\rm coll}$ is an effective occupation number. We hence obtain $\Gamma_{\rm coll}/J_N\sim{\rm const}$ and conclude that decays induced by the OMTs do not limit the number of nodes, provided we are close enough to the mode (s,0). In contrast, when tuning to the anti-symmetric modes, there are always $N-1$ decay channels present, leading to $\Gamma/J_N \propto N$. Finally, note that even for the symmetric modes there are always the intrinsic decoherence processes of the qubits that set an absolute time-scale and limit the number of nodes via the requirement $J_N T_2\propto J T_2/N \gg 1$.

\section{Implementations}
\label{sec:realizations}

\begin{figure}
\includegraphics[width=0.5\textwidth]{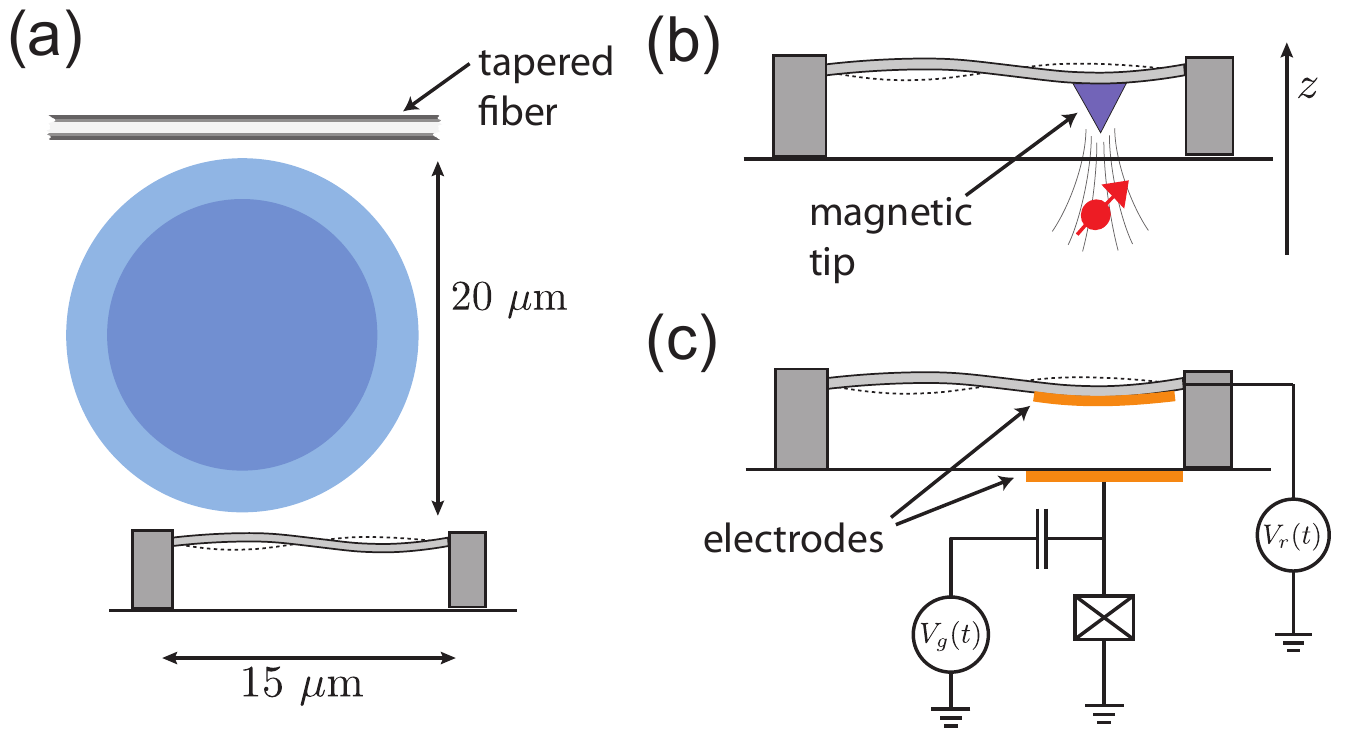}
\caption{(Color online) Realization with toroidal cavities and doubly clamped beams. 
(a) Illustration of a single toroidal cavity coupled to an optical fiber as well as to a doubly clamped beam. 
(b) Coupling of the doubly clamped beam to a spin qubit via magnetic fields. 
(c) Coupling of the doubly clamped beam to a charge qubit via electrostatic forces. }
\label{fig:realizations}
\end{figure}

In the preceeding sections, we have mainly discussed the OMT on general grounds and conditions for low-noise operation have been given. Here, we describe the realization of the OMT within specific physical systems and the resulting numbers have already been used in \secref{sec:longDistance} and \ref{sec:onchip} to illustrate the performance of the proposed schemes. Generally, implementing the OMT poses the challenge of combining high-$Q$ optical cavities with high-$Q$ mechanical resonators exhibiting an appreciable zero-point motion, i.e., a low effective mass. Further, the mechanical resonators must interact with the qubits, which requires suitable coupling schemes. Linking the different nodes of a quantum network can be achieved with standard optical fibers, while for short distances, integrated  nano-photonic circuits \cite{Safavi-Naeini2011,Reed2004} might be a promising alternative, which we, however, do not explore here.

\subsection{Toroidal cavities and doubly clamped beams}

As a possible candidate system for the OMT we propose to use a setup similar to the one recently demonstrated by Anetsberger \etal{} \cite{Anetsberger2009}. There, a doubly clamped beam of stressed SiN is positioned in the evanescent field of a whispering-gallery mode (WGM) supported by a microtoroidal cavity made of silica. This arrangement gives rise to the standard OM coupling and the toroid is further coupled to a tapered fiber for driving and interrogation as shown in \figref{fig:realizations}(a). To obtain a resonator-qubit coupling  we propose to attach  a coupling element to the resonator (e.g. a magnetic tip or an electrode as discussed below), which resides sufficiently far outside the evanescent field of the WGM such that the optical quality factor is not degraded. To be specific, we envision to use a toroid of diameter $2R\approx20\unit{$\mu$m}$ and the second harmonic mode of a doubly clamped beam of length $l\approx15\unit{$\mu$m}$ positioned as in \figref{fig:realizations}(a). The coupling to qubit and cavity can then occur at different antinodes and the coupling element is located at a distance of $d\gtrsim2.5\unit{$\mu$m}$ from the rim of the toroid, which is sufficient for our purposes (see below).

The elastic modes of a doubly-clamped SiN beam of length $l$, width $w$ and thickness $t$ are well described by standard elastic theory \cite{Unterreithmeier2010, Anetsberger2009}. For the case that the beam is exposed to high tensile stress, the modes resemble those of a vibrating string with frequencies given by $\om_n=\sqrt{\sigma_0/\rho_m} \,\pi n/l$, where $n=1,2,\ldots$ labels the mode, $\rho_m$ is the mass density and $\sigma_0$ the internal tensile stress. In contrast, for low internal tensile stress and $w=t$ the frequencies scale as $\om_n=c^2_n \sqrt{E/12\rho_m}\, t/l^2$ with $c_1\approx 4.7300$, $c_2\approx 7.8532$ \cite{Cleland}, where $E$ is Young's modulus. 
We are interested in the second harmonic of the beam and hence set $\om_r=\om_2$. When quantizing the resonator, the effective mass of each mode is given by an integral over the mode's displacement profile $u_n(z)$ according to $m_\eff=m l^{-1} \int_0^l \rmd z\,u_n^2(z)$, where $m=t^2l\rho_m$ is the real mass. For the relevant mode we find that $m_\eff\approx m/2$ in both stress regimes. By tuning the internal tensile stress, one can interpolate between the two regimes and values of $\rho_m=2800\unit{kg/m${}^2$}$, $E=160\unit{GPa}$, and $\sigma_0$ up to $\sim 1\unit{GPa}$ have been demonstrated (see Refs.\,\cite{Unterreithmeier2010, Anetsberger2009}). Therefore, 
$\om_r/2\pi\approx 5\ldots50\unit{MHz}$ for a beam with dimensions $(l,w,t)=(15,0.05,0.05)\unit{$\mu$m}$ is realistic and the respective zero-point motions are in the range $a_0=\sqrt{\hbar/2 m_\eff \om_r}\approx 1.8\ldots0.6\times 10^{-13}\unit{m}$.
Demonstrated $Q$-factors for these devices range from $10^4$ to $1.5 \times 10^5$ \cite{Anetsberger2009, Unterreithmeier2010} and for $Q=2\times 10^5$ we obtain, e.g., a mechanical decoherence rate of $\gamma_m N_m/2\pi\approx 10\unit{kHz}$ at a support temperature of $T\approx100\unit{mK}$.

The microtoroidal cavity is modeled by a pair of (by symmetry) degenerate counter-propagating WGM modes, whose field distributions we denote by $E_{L,R}(\vecc{r})$. The theoretical quality factors (including radiation loss and absorption in silica)  are beyond $10^{10}$, whereas demonstrated values reach up to $Q=4\times 10^8$ \cite{Spillane2007}. When the SiN beam is brought into the near-field of these modes, they generally experience a frequency shift as well as a cross-coupling.
Simple perturbation theory on the level of the wave equation shows that the relevant overlap integrals are $I_{\mu\nu}\propto\int_{\rm resonator} \rmd^3 r\, \Delta\epsilon\, E^*_\mu E_\nu$ with $\Delta\epsilon=n_{\rm res}^2-1$ and $n_{\rm res}$ the refractive index of the resonator. However, a resonator oriented as shown in \figref{fig:realizations}(a) is quite different from a point-scatterer and hence does not scatter between the modes. To be specific, we obtain the scaling $I_{LR}/I_{RR}\sim \exp[-k^2R/\chi]\ll 1$, where $k=2\pi/\lambda_c$ and $\chi^{-1}=250\unit{nm}$ is the field decay constant outside the cavity \cite{Anetsberger2009}. Therefore, we can take the OM coupling to be of the type assumed in \eeqref{eq:Hom} with $g_0 \propto \partial I_{RR} /\partial z$.

We now turn to linking the different nodes:
When the toroidal cavities are side-coupled to the fiber, the modes of a given propagation direction experience a uni-directional coupling, as needed for the cascaded quantum network of \secref{sec:longDistance}. However, for standing wave modes, as they occur when the degeneracy of the WGM-modes is lifted by an amount $h\gtrsim\kappa,G$ or alternatively in Fabry-P\'erot or photonic band-gap cavities, the cascaded coupling has to be achieved by additional non-reciprocal optical elements, such as optical circulators or Faraday rotators. While this may complicate the structure of the network, the idea of the OMT remains valid.

In the on-chip setting, the nodes are linked by a cavity of length $L$ and refractive index $n_{\rm fib}$, such that its mode-spacing is given by $\delta\om=\pi n_{\rm fib}c/L$. The single-mode description of \secref{sec:onchip} is valid as long as $\delta\om$ is much larger than all other frequency scales, which poses the restriction $L\lesssim 10\unit{cm}$ for mechanical frequencies of $\om_r/2\pi\lesssim100\unit{MHz}$.
Such a cavity could, e.g., be realized by terminating an optical fiber with coated end mirrors. For the mirrors demonstrated in Ref.\,\cite{Hunger2010} (amplitude loss of $85\unit{ppm}$) we obtain the naive estimate $\kappa_{0f}/2\pi\sim 10\unit{kHz}$ for the decay rate of a cavity with $L=10\unit{cm}$ and assuming a power-attenuation of $1\unit{dB/km}$, the intrinsic losses are of the same order.

\subsection{Spin and charge qubits}

In principle, the proposed OMT works for all basic types of qubits that can be coherently coupled to a mechanical resonator. In the following, we discuss two prominent examples.

\subsubsection{Spin qubits}

Prime examples for spin qubits \cite{Morton2011} are phosphor donors in silicon \cite{Tyryshkin2003} or the ground state triplet of NV centers in diamond \cite{WrachtrupJelezko2006}.
Both systems exhibit excellent coherence properties, i.e., $T_2=1.4\unit{ms}$ for NV-centers \cite{Balasubramanian2009} and $T_2=10\unit{ms}$ for phosphor donors \cite{Tyryshkin2003}, with more recent experiments even achieving $T_2$ on the order of seconds \cite{Tyryshkin2011}. Also, coherent control of both systems via external fields has been demonstrated. 

The coupling to the mechanical resonator is accomplished by attaching a magnetic tip to the beam at the position of an antinode as shown in \figref{fig:realizations}(b). This tip produces a magnetic field at the location of the spin which depends on the resonator displacement, and to first order in the latter we obtain
\begin{align}
H_q=\frac{1}{2}\vec{B}_0\cdot\vec{\sigma}+\frac{\lambda}{2}(b+b^\dagger)\sigma_z \,,
\end{align}
where $\vec\sigma$ is the vector of Pauli matrices, while $\vec{B}_0$ is the free qubit splitting, into which the constant component of the tip's field has been absorbed. The qubit-resonator coupling is  given by 
$\lambda=g_s \mu_B a_0 \abs{\nabla B}/\hbar$, 
with $g_s\approx 2$ being the gyro-magnetic ratio,  $\mu_b$  the Bohr magneton, and 
$\abs{\nabla B}$ the field gradient produced by the tip at the location of the qubit.
Provided that the qubit splitting can be arranged to be $\vec{B}_0=(\om_q,0,0)$ with $\om_q\approx\om_r$, we may store the qubit in the $\sigma_x$-eigenstates and after a relabeling of axes ($x\leftrightarrow z$) and subsequent RWA directly obtain the coupling of \eeqref{eq:Hnode}, which was the starting point for our analysis.

For qubits whose natural energy splitting $\om_0$ does not match the resonator frequency (i.e. $\om_0\gg \om_r$), we can apply a classical drive of frequency $\om_d$ to bridge the gap between the two energy scales. In this case we take $\vec{B}_0(t)=(2\Omega\cos(\om_d t),0,\om_0)$, where $\Omega$ is the Rabi-frequency.
After moving to a frame rotating at $\om_d$ we drop rapidly oscillating terms and transform to a new eigenbasis basis rotated by an angle $\theta$. The latter is given by $\tan\theta=\Omega/\Delta$, where $\Delta=\om_0-\om_d$ is the detuning of the qubit from the drive. The qubit Hamiltonian then becomes 
\begin{align}
H_{q}=\frac{\om_q}{2} \sigma_z+\frac{\lambda}{2} (b+b^\dagger)\left( \cos(\theta) \sigma_z - \sin(\theta)\sigma_x\right)\,,
\end{align}
with the effective energy splitting $\om_q=\sqrt{\Delta^2+\Omega^2}\ll\om_0$.
Provided that we have $\lambda\ll\om_q\sim\om_r$, we may drop all rapidly oscillating terms in a RWA and retain only 
\begin{align}
H_{\rm{q-res}}=\frac{\lambda}{2}\sin(\theta) \left( \sigma^- b^\dagger + \sigma^+ b \right)\,,
\end{align}
which is of the form given in \eeqref{eq:Hnode} if we redefine $\sin(\theta)\lambda\rightarrow\lambda$.

We take the magnetic tip coupling the beam to the qubit to be fabricated of highly magnetic Co$_{70}$Fe$_{30}$ which exhibits a magnetization of $M=2.3\unit{T/$\mu_0$}$ \cite{Mamin2007}.  For a cone-shaped tip of height and radius $100\unit{nm}$ we estimate the field gradient $25\unit{nm}$ below the tip to be $\abs{\nabla B}=10^7\unit{T/m}$. Using a resonator of low frequency $\om_r/2\pi\approx 5\unit{MHz}$ and thus high zero-point motion $a_0\approx 1.8\times 10^{-13}\unit{m}$, we obtain a coupling of $\lambda/2\pi\approx50\unit{kHz}$.
Finally, we estimate the degradation of the toroid's quality factor due to the presence of the magnetic tip, which is placed a distance $d\gtrsim2.5\unit{$\mu$m}$ away from the rim of the toroid as discussed above. We make the conservative assumptions that all light which hits the tip gets lost and obtain
$Q\gtrsim 2\pi e^{2\chi d} V_{\rm mode}/ \lambda A_{\rm tip} $.
Here, $V_{\rm mode}=\int\rmd^3 r \,\epsilon_r(r)\, \abs{E(r)}^2/\abs{E_{\rm max}}^2$ is the mode volume of the WGM, $A_{\rm tip}$ the cross-sectional area of the tip and $\chi$ the decay constant of the evanescent cavity field. Using the values $A_{\rm tip}\approx 10^{-2}\unit{$\mu$m$^2$}$, $V_{\rm mode}=90\unit{$\mu$m$^3$}$ \cite{Spillane2007}, $\chi^{-1}=250\unit{nm}$ \cite{Anetsberger2009} we obtain $Q\gtrsim 10^{13}$ which is well above the intrinsic $Q$ quoted above.

Based on these numbers, the results presented in \secref{sec:stateTransfer} and \secref{sec:onchipFidelity} show that quantum networking operations for spin qubits are feasible, both in the long- and short-distance setups.

\subsubsection{Charge qubits}

A second promising qubit candidate for the setup described in this work is the superconducting charge qubit.  It has evolved from the original Cooper pair box qubit \cite{Makhlin2001} to more recent designs such as the transmon \cite{Koch2007,Schreier2008}. For the latter,
coherence times of $T_2=2\unit{$\mu$s}$ have been observed \cite{houck2009} and recently, the relaxation times have been pushed up to $T_1=200\unit{$\mu$s}$ \cite{Kim2011}. Control or measurement tasks can, for example, be performed in a circuit QED architecture \cite{Mallet2009}.

Coupling a charge-based qubit to a mechanical resonator is most easily achieved by pushing excess charges onto an electrode located on the resonator as depicted in \figref{fig:realizations}(c). The resulting coupling amounts to a position-dependent capacitance and has been widely discussed and also implemented in experiment \cite{Armour2002,IrishSchwab2003,Martin2004,LaHaye2009}. However, since the intrinsic level splitting $\om_0$ of superconducting qubits is typically in the GHz range, we need to bridge the gap to the MHz resonator frequency $\om_r$ in order to obtain a Jaynes-Cummings coupling. Therefore, we modulate the gate voltages as described in Ref.\,\cite{Martin2004,Rabl2004} with frequency $\om_d=\om_0-\om_r$ and amplitude $V_r^0$, which yields the coupling of \eeqref{eq:Hnode} with 
\begin{align}
\lambda=\frac{2}{\hbar} E_C \frac{a_0}{d_0}\frac{C_r V_r^0}{2 e}\,.
\end{align}
Here, $E_C$ is the charging energy of the qubit, $C_r$ the equilibrium resonator capacitance, and $d_0$ the equilibrium electrode separation of the coupling capacitance. For $E_C/h\approx20\unit{GHz}$, $d_0\approx100\unit{nm}$, and $V_r^0$ a couple of volts one easily obtains values in the MHz regime (see also Refs.\,\cite{IrishSchwab2003,LaHaye2009}). If it is necessary to place the qubit further away from the resonator, the coupling could also be mediated by an intermediate normal-conducting wire, or additional superconducting circuitry \cite{Regal2011}.

We estimate the optical $Q$-factor due to the presence of the electrode, which we assume to cover one third of the resonator and to be thinner than the typical skin-depth of $\sim10$nm. The field then penetrates the whole object and an estimate for the $Q$-factor is given by  $Q\gtrsim e^{2\chi d}V_{\rm mode}/(V_{\rm electrode}\epsilon^\prime_r)$ \cite{Harrington}, where $\epsilon_r^\prime$ is the imaginary part of the relative permittivity of the tip and $V_{\rm electrode}=2.5\times 10^{-3}\unit{$\mu$m$^{3}$}$ its volume. Using  $\epsilon_r^\prime<10$ for noble metals at optical frequencies \cite{JohnsonChristy1972} we find $Q\gtrsim 10^{12}$, which again exceeds the intrinsic $Q$.

As has been discussed in \secref{sec:longDistance} and \ref{sec:onchip}, the fidelities for quantum networking operations resulting from the above numbers are mainly limited by intrinsic qubit-dephasing. Improvements of $T_2$ would thus have a big positive impact on the performance and also, especially superconducting qubits would profit from leaving the regime where the evolution of the OM degrees of freedom is adiabatic.

\section{Conclusions \& Outlook}
\label{sec:conclusion}

In conclusion, we have discussed the potential of micro-mechanical resonators to mediate interactions between dark qubits and light. We have described in detail the dynamics of a single OMT and found that the conditions for low-noise operation overlap with those for OM ground-state cooling. Further, we have derived an effective description of a multiqubit cascaded network that is suitable for long-distance quantum communication, discussed the relevant noise-sources in such systems, and presented an efficient protocol for state transfer between two nodes. In addition, we have shown how OMTs can be utilized in an on-chip setting to perform entangling gates between qubits that are spaced by less than $\sim 10\unit{cm}$. Here, OMTs might provide an alternative to other coupling schemes
\cite{Bialczak2011,Majer2007,Kielpinski2002,Taylor2005,Rabl2010} 
for scalable quantum processors. 

We point out that the ingredients needed for the OMT are quite generic and may be implemented with a variety of solid-state qubits and different OM systems in the optical, as well as the microwave domain \cite{Regal2011,Teufel2011}. In this work, we have specifically discussed a potential realization for spin and charge qubits, showing that long- and short-distance communication is feasible for present-day technology.
The gate-schemes presented in this work only provide a starting point for optimized and more complex protocols. In particular, the long-distance state transfer protocol could be extended by schemes to correct for photon loss (see, e.g., Ref.\,\cite{vanEnk1997}), while the on-chip gates could profit from spin-echo techniques to reduce the impact of qubit dephasing. In both settings, the limitations posed by imperfect qubits could be overcome by departing from the adiabtic desciption of the OMT and devising fully dynamical gate sequences. This seems especially worthwhile for charge qubits, where the qubit-resonator coupling $\lambda$ can be made large rather easily. The benefit would be shorter gate-sequences at the price of possibly stronger requirements on controllability.

More generally, the OMT can be used to realize complex qubit-light interactions, which may have applications in the implementation of nonlinear optical devices on a few-photon level.  These are particularly interesting in the context of integrated nanophotonic systems, where the proposed device naturally fits in.

\section*{Acknowledgments} We gratefully acknowledge discussions with Klemens Hammerer. This work is supported by ITAMP, NSF, CUA, DARPA, the Packard Foundation, and the Danish National Research Foundation. Work in Innsbruck is supported by the Austrian Science Fund (FWF) through SFB FOQUS and the EU network AQUTE.

\appendix

\section{Elimination of the optomechanical system}
\label{sec:appGeneralElimination}

We briefly describe the elimination of the OM degrees of freedom to obtain an effective description of the qubit dynamics. We denote by $\rho$ the density matrix of the complete $N$-node system composed of qubit and OM degrees of freedom and our goal is to derive an effective ME for the reduced qubit density operator $\mu=\tr_{\rm om}\{\rho\}$ to second order in the qubit-resonator coupling $\lambda$ by means of projection operator techniques \cite{BreuerPetruccione2002}. The way the various OM systems interact with each other is left unspecified in this Appendix and we lump the whole OM dynamics into a Liouvillian $\mL_{\rm om}$ equivalent to the linearized OM QLEs of the setup under consideration. In an interaction picture with respect to the free qubit Hamiltonian $\om_q\sum_i\sigma_z^i/2$ the ME for the complete setup then reads:
\begin{align}
\label{eq:MEintpic}
\dot\rho &= \mL(t)\rho \equiv \left(\mL_{\rm om}+\mL_{\rm int}(t)\right)\rho\,, \\
\mL_{\rm int}(t)\rho&= -i \frac{\lambda}{2}\sum_i \left[ b_i \sigma^+_i e^{i\om_q t} + b_i^\dagger \sigma^-_i e^{-i\om_q t} , \rho \right]\,,
\end{align}
where $\mL_{\rm int}$ describes the qubit-resonator interactions. Note that $\sigma_i^-$, $b_i$ can be nodal operators (\secref{sec:longDistance} and \appref{sec:appLongFiber}) or normal mode operators (\secref{sec:onchip} and \appref{sec:appShortFiber}) and that the commutators of the $\sigma_i^-,\sigma_i^+$ will not enter from now on. 
We define the projector on the relevant part of the density matrix as
$\mP\rho=\tr_{\rm om}\{\rho\}\otimes \rss$, where $\rss$ is the steady state of the OM system in the absence of the qubits, defined by $\mL_{\rm om}\rss=0$. We also introduce $\mQ=1-\mP$ and note that $\mP^2=\mP$, $\mQ^2=\mQ$ and $\mP\mQ=\mQ\mP=0$. To derive an effective equation of motion for $\mu$ we project \eeqref{eq:MEintpic} on the $\mP$- and $\mQ$-subspaces and formally integrate the equation for the $\mQ$-part, where we assume for simplicity that at the initial time $\mQ\rho(t_0)=0$. The formal expression for the $\mP$-part is then
\begin{align*}
&\mP\dot\rho = \mP\mL(t)\mP\rho \\
&+ \mP\mL(t)\mQ\int_{t_0}^t \rmd s\, T\!\exp\left[\int_{s}^{t}\rmd\tau\,\mQ\mL(\tau)\mQ \right] \mQ \mL(s)\mP\rho(s) \,,
\end{align*}
where $T\!\exp[\ldots]$ is the time-ordered exponential. 
Considerable simplification is brought about by exploiting the relations
\begin{align*}
\mL_{\rm om}\mP &= 0\qquad (\textrm{steady state})\,,\\
\mP\mL_{\rm om} &= 0\qquad (\mL_{\rm om}\textrm{ preserves trace})\,,\\
\mP\mL_{\rm int}\mP\rho &= 0 \qquad (\textrm{vanishing of }\tr_{\rm om}\{\rss b_i\}=\smean{b_i}_{\rm free})\,,
\end{align*}
where the last conditions follow from the fact that we have removed any classical forces from the description of the OM systems.
These relations allow us to drop $\mL_{\rm om}$ everywhere except for the exponent and since we are interested in the dynamics to second order in $\lambda$ we may subsequently drop $\mL_{\rm int}$ in the exponent leaving us with
\begin{align}
\label{eq:finalExp}
  \mP\dot\rho(t)
&=\mP \mL_{\rm int}(t)\int_0^{t-t_0}\rmd\tau \, e^{\mL_{\rm om}\tau} \mL_{\rm int}(t-\tau)\mP \rho(t-\tau) \,.
\end{align}
Upon expanding the interaction Liouvillian and taking the trace implicit in the left-most $\mP$-operation we obtain terms of the following form (neglecting transients by sending $t_0\rightarrow -\infty$)
\begin{align*}
\dot{\mu} \propto \lambda^2\int_0^{\infty}\rmd\tau\, e^{\pm i \om_q \tau } f(\tau) \mL_A \mL_B \mu(t-\tau)\,,
\end{align*}
where $f(\tau)=\smean{A(\tau)B(0)}_{\rm free}=\tr_{\rm om}\{Ae^{\mL_{\rm om}\tau}B\rss\}$ is a resonator-resonator correlation function  evaluated in the steady state of the OM system and $\mL_{A,B}$ are associated qubit Liouvillians. For illustration, we assume that $f(\tau)$ has a dominant contribution $f(\tau)\propto e^{(i\Omega-\gamma)\tau}$ and with $\epsilon=i(\Omega\pm\om_q)-\gamma$ integration by parts then yields
\begin{align*}
\dot{\mu} \propto \frac{\lambda^2}{\epsilon}\mL_A \mL_B \mu(t)  - \frac{\lambda^2}{\epsilon} \int_0^{\infty}\rmd\tau\, e^{\epsilon\tau} \mL_A \mL_B  \dot\mu(t-\tau) \,.
\end{align*}
By iterating this equation we obtain an expansion in $\lambda/\abs{\epsilon}$ and for $\lambda\ll\abs{\epsilon}$ the first, Markovian term dominates. If this is true for all terms in \eeqref{eq:finalExp} and for all frequency components of the correlation functions, we may neglect the non-Markovian  corrections, which is equivalent to replacing $\rho(t-\tau)\rightarrow\rho(t)$ in \eeqref{eq:finalExp}. In general, this procedure is thus valid if the OM normal modes which couple to the qubits decay much faster than $\lambda^{-1}$ or if they are detuned from the qubits by much more than $\lambda$. 
The final result can be compactly written as a Markovian ME with coefficients given by one-sided Fourier transforms of resonator correlation functions evaluated at the qubit frequency:
\begin{align}
\label{eq:effME}
\dot\mu=-\frac{\lambda^2}{4}\sum_{i,j}\Big[
      &S_{ij}(\om_q)(\sigma_i^+\sigma_j^-\mu - \sigma_j^-\mu\sigma_i^+)\\
   + &T_{ij}(\om_q)(\sigma_i^-\sigma_j^+\mu - \sigma_j^+\mu\sigma_i^-)
   +\hc \Big] \,,\nonumber 
\end{align}
where we have dropped terms rotating at $\exp[\pm i2\om_q t]$ that contain operators such as $\sigma_i^+\sigma_j^+$ based on the assumption $\lambda\ll\om_q$. The coefficients are given by
\begin{align}
\label{eq:corrS}
S_{ij}(\om)&=\int_0^\infty\rmd\tau\,\smean{b_i(\tau)b_j^\dagger(0)}_{\rm free} \,e^{i\om\tau}\,,\\
T_{ij}(\om)&=\int_0^\infty\rmd\tau\,\smean{b_i^\dagger(\tau)b_j(0)}_{\rm free} \,e^{-i\om\tau}\,,
\end{align}
and for later convenience we also introduce $X_{ij}(\om)=S_{ij}(\om)-T^*_{ij}(\om)$ and $Y_{ij}(\om)=T_{ij}(\om)+T^*_{ji}(\om)$ such that
\begin{align}
\label{eq:corrX}
X_{ij}(\om)&=\int_0^\infty\rmd\tau\,\smean{[b_i(\tau),b_j^\dagger(0)]}_{\rm free} \,e^{i\om\tau}\,,\\
\label{eq:corrY}
Y_{ij}(\om)&=\int_{-\infty}^\infty \rmd\tau \smean{b_i^\dagger(\tau)b_j(0)}_{\rm free} \,e^{-i\om\tau}\,.
\end{align}
It will turn out that interaction- and decay-rates are given by $X_{ij}(\om)$, while $Y_{ij}(\om)$ occurs in diffusion terms. 

\section{Master equation for cascaded setup}
\label{sec:appLongFiber}

Here, we derive the effective ME for $N$ qubits coupled to a chain of cascaded OM systems as discussed in \secref{sec:longDistance}. Also, for $N=1$ we reproduce the results of \secref{sec:singleNode} in a different language. To begin with, we rewrite the general result \eqref{eq:effME} to separate single- and multiqubit terms (this rewrite is exact): 
\begin{align}
\label{eq:effCascMEapp}
\dot\mu 
&= \sum_{i}\left\{ 
-i \frac{ \Delta_i}{2}\left[\sigma_z^i,\mu\right]\right. \nonumber\\
&\phantom{=}+ \left. \frac{\Gamma_i}{2}(N_i+1)\,\mD[\sigma_i^-]\mu
 +\frac{\Gamma_i}{2} N_i\,\mD[\sigma_i^+]\mu 
\right\} \nonumber\\
&\phantom{=}- \sum_{i\neq j}\left(J_{ij} \left[\sigma_i^+,\sigma_j^-\mu\right]+J_{ij}^*\left[\mu\sigma_j^+,\sigma_i^-\right]\right)\nonumber\\
&\phantom{=}+ \sum_{i\neq j} D_{ij} \left[\left[\sigma_j^+,\mu\right],\sigma_i^-\right] \,,
\end{align}
where $\mD[a]\mu=2a\mu a^\dag - a^\dag a\, \mu - \mu\, a^\dag a$ denotes a Lindblad term with jump operator $a$. Here, we have defined the effective frequency shifts $\Delta_i=\Delta_{0,i} + \Delta_{{\rm th},i}$ with contributions
\begin{align}
\Delta_{0,i}&=\frac{\lambda^2}{4}\im\left\{ X_{ii}(\om_q)\right\}\,, &
\Delta_{{\rm th},i}&=-\frac{\lambda^2}{2}\im\left\{ T_{ii}(\om_q)\right\}\,,
\end{align}
as well as the effective decay rates $\Gamma_i$ and occupation numbers $N_i$, 
\begin{align}
\Gamma_i&=\frac{\lambda^2}{2}\re\left\{ X_{ii}(\om_q)\right\}\,, &
N_i&=\frac{\lambda^2}{4\Gamma_i}Y_{ii}(\om_q)\, .
\end{align}
Further, the multiqubit coupling and diffusion rates are given by
\begin{align}
\label{eq:multiQubitRates}
J_{ij}&=\frac{\lambda^2}{4} X_{ij}(\om_q)\,,\qquad
D_{ij} = \frac{\lambda^2}{4} Y_{ij}(\om_q)\,,
\end{align}
respectively, 
and we stress that  the cascaded nature of the coupling will only become evident below, where we evaluate $J_{ij}$ from the physics of the underlying, eliminated system.

The OM steady state correlation functions \eqref{eq:corrS}-\eqref{eq:corrY} determining the effective dynamics are conveniently calculated from the QLEs\,\eqref{eq:linQLEcasc} and the input-output relation \eqref{eq:cascInOut} by Fourier or Laplace transformation. To give the resulting expressions in a compact form we use the OM response matrices $A^i(\om)=(M^i-i\om \mathbb{1})^{-1}$ and in addition, we define the single-node transfer matrices $C^i(\om)$ that describe how the fiber-field is modified when bypassing node $i$. They are given by  $C^i(\om)=(P-2\kappa_f^i PA^i(\om) P)$, where $P={\rm diag}[0,1,0,1]$. All relevant quantities can then be written in terms of the multinode response-matrices $\mT^{ij}(\om)$ defined by ($i>j>0$)
\begin{align*}
\mT^{jj}(\om)&=A^j(\om)\,,\\
\mT^{ij}(\om)&=-2\sqrt{\kappa_{\smf}^{i}\kappa_{\smf}^{j}}A^i(\om)C^{i-1}(\om)\cdots C^{j+1}(\om) P A^j(\om)\,.\nonumber
\end{align*}
In addition, to aid the description of the common fiber input, we also introduce ($i>0$)
\begin{align}
\mT^{i0}(\om)&=-\sqrt{2\kappa_\smf^i}A^i(\om)C^{i-1}(\om)\cdots C^{1}(\om)\,.
\end{align}

To determine the correlation functions $X_{ij}(\om)$ we make use of the quantum regression theorem \cite{QuantumNoise} with the initial condition $\smean{[b_i(0),b_j^\dagger(0)]}_{\rm free}=\delta_{ij}$ and solve the resulting equations in the Laplace domain. 
Since node $i$ is only driven by nodes $j<i$, we obtain
\begin{align}
X_{ij}(\om) = \begin{cases}
0                  & \text{for  $i < j$} \\
\mT^{ij}_{11}(\om) & \text{for  $i \ge j$} 
\end{cases}
\end{align}
which, via \eeqref{eq:multiQubitRates}, turns \eeqref{eq:effCascMEapp} into a cascaded ME.

In order to specify the noise terms in the ME we solve \eeqref{eq:linQLEcasc} in the frequency domain, with the Fourier transformation of some quantity $f(t)$ defined as $f(\om)=\frac{1}{\sqrt{2\pi}}\int\rmd t \,e^{i\om t}f(t)$ (which induces $[f(\om)]^\dagger=f^\dagger(-\om)$). As a result, we get $\vecc{v}^i(\om)=\mT^{i0}(\om)\vecc{I}(\om) -\sum_{j\leq i} \mT^{ij}(\om)\vecc{R}^j(\om)$, where $\vecc{I}(\om)\equiv\vecc{I}^1(\om)=(0,f_{\rm in}(\om),0,f_{\rm in}(\om))^T$ is the fiber input of the first node. Due to the $\delta$-correlated nature of the noise sources we can express the second moments of the OM operators as follows ($i,j=1\ldots N$ label nodes and $k,l=1\ldots4$ vector components):
\begin{align}
\smean{v^{i\dagger}_k(\om^\prime) v^j_l(\om) }_{\rm free}=\mC^{ij}_{kl}(\om)\delta(\om^\prime+\om)\,,
\end{align}
Here, we have introduced the matrices $\mC^{ij}(\om)$, which directly give the desired correlators for the ME via $Y_{ij}(\om)=\mC^{ij}_{11}(\om)$. 
They can be expressed in terms of the transfer matrices and noise statistics according to
\begin{align}
\mC^{ij}(\om)&=(\mT^{i0}(\om))^*I^0 (\mT^{j0}(\om))^T \nonumber\\
&+\sum_{n=1}^{{\rm min}(i,j)}(\mT^{in}(\om))^*r^n(\mT^{jn}(\om))^T\,,
\end{align}
where the matrices $r^i$ characterize the noise correlations of node $i$ by means of  $\smean{R^{i\dagger}_k(t) R^{i}_l(t^\prime)}=r^i_{kl}\delta(t-t^\prime)$, while $I^0$  describes the fiber input, $\smean{I^{\dagger}_k(t) I_l(t^\prime)}=I^0_{kl}\delta(t-t^\prime)$. From the statistics quoted in the main text it follows that $r^i={\rm diag}[\gamma_m^i N_m^i,0,\gamma_m^i (N_m^i+1),2\kappa_0]$ and $I^0={\rm diag}[0,0,0,1]$.

Finally, we have to evaluate the thermal frequency shift $\Delta_{{\rm th},i}$, which we do exemplarily for the first node. Using the quantum regression theorem we obtain the relevant correlation function $T_{11}(\om)=\sum_k A^{1*}_{1k}(\om) \smean{v_k^{1\dagger} v^1_1}_{\rm free}$, where the steady state moments are given by the frequency integral $\smean{v_k^{i\dagger} v^i_l}_{\rm free}=\int\frac{\rmd\om}{2\pi}\,\mC_{kl}^{ii}(\om)$, which can be evaluated exactly
\cite{Gradshteyn2007}.
If the qubit is on resonance with a normal mode of the OMT we obtain for $\Delta_c=\om_r$
\begin{align}
\frac{\Delta_{\rm th}}{\Delta_0}\approx
\frac{\gamma_m N_m}{G^2/\kappa}
+\frac{\kappa^2}{2\om_r^2}\,,
\end{align}
where we have neglected higher orders in $\kappa/\om_r$, $G/\om_r$ as well as non-RWA corrections 
to the thermal part. From the low noise conditions identified in \secref{sec:singleNodeImperfections} it follows that $\abs{\Delta_{\rm th}}\ll\abs{\Delta_0}$. If, however, the qubit is off-resonant with the OM system as for $G\rightarrow 0$ on the trajectory of \figref{fig:singleNode}(c), we find that the leading orders are still given by the above expression. 
In conclusion, we thus have $\abs{\Delta_{\rm th}}\lesssim\abs{\Delta_{0}}$ for most cases of interest and finally note that $\Delta_{\rm th}$ can be compensated to the same amount as the unavoidable shift $\Delta_0$ by tuning of the bare qubit frequencies.

\section{Master equation for setup with fiber cavity}
\label{sec:appShortFiber}

Here, we derive the effective qubit dynamics for the case of the on-chip setup discussed in \secref{sec:onchip}. Note that the results given here apply to the case of identical nodes.

\subsection{Network with $N$ identical nodes}

According to the second term in \eeqref{eq:Hosc}, only the "center-of-mass" mode couples to the fiber cavity and we therefore introduce normal modes for the cavities according to $\tilde c_i = V_{ij} c_j$ with $V_{1j}=1/\sqrt{N}$, and similarly for mechanical resonators and qubits. The remaining rows of $V$ are constructed from the constraint $V^TV=\mathbb{1}$, such that the remaining terms in Eqs.\,\eqref{eq:Hlin},\eqref{eq:Hosc} are left invariant. As a result, we can write the QLEs describing the OM variables as $N$ decoupled sets
\begin{align}
\label{eq:QLEapp}
\dot{ \ti{\vecc{v}}}^i(t) = -\tilde M^i \ti{\vecc{v}}^i(t) - \ti{\vecc{R}}^i(t)\,, 
\quad i=1\ldots N\,,
\end{align}
where $\ti{\vecc{v}}^1=(\ti b_1,\ti c_1, d_0,\ti b_1^\dag,\ti c_1^\dag,d_0^\dag)^T$ and $\ti{\vecc{v}}^{i\ge 2}=(\ti b_i,\ti c_i,\ti b_i^\dag,\ti c_i^\dag)^T$. Note that the systems $i\ge2$ are all identical, which will simplify the final results. The noise vectors $\ti{\vecc{R}}^i(t)$ can be constructed easily from the original QLEs \eqref{eq:onchipQLEs} and the orthogonality of the transformation $V$ ensures that they are mutually uncorrelated. 
We have
$\ti{\vecc{R}}^1(t)=(\sqrt{\gamma_m}\tilde \xi_1,\sqrt{2\kappa_0}\tilde f_{0,1},\sqrt{2\kappa_{0f}} f_{0},\sqrt{\gamma_m}\tilde \xi^\dagger_1,\sqrt{2\kappa_0}\tilde f^\dagger_{0,1},\sqrt{2\kappa_{0f}} f_{0}^\dagger)^T$ 
and 
$\ti{\vecc{R}}^{i\ge2}(t)=(\sqrt{\gamma_m}\tilde \xi_i,\sqrt{2\kappa_0}\tilde f_{0,i},\sqrt{\gamma_m}\tilde \xi^\dagger_i,\sqrt{2\kappa_0}\tilde f^\dagger_{0,i})^T$. 
The statistics of these operators are given by 
$\smean{(\ti{R}^i_k(t))^\dag \ti{R}^i_l(t^\prime)}=\tilde r^i_{kl}\delta(t-t^\prime)$
with diagonal correlation matrices 
$\ti{r}^1={\rm diag}[\gamma_m N_m ,0,0,\gamma_m (N_m+1),2\kappa_0,2\kappa_{0f}]$ 
and 
$\ti{r}^{i\ge2}={\rm diag}[\gamma_m N_m ,0,\gamma_m (N_m+1),2\kappa_0]$.
Finally, the drift  matrices of the  systems \eqref{eq:QLEapp} are $\tilde M^{i\ge 2}=M$, where $M$ is defined in \eeqref{eq:OMmatrix} with $\kappa\rightarrow\kappa_0$ and 
\begin{align*}
\tilde M^1=
i \begin{pmatrix}
\om_r      & G^*        & 0           & 0          & \zeta G   & 0     \\
G          & \Delta_c   & K^*         & \zeta G    & 0         & 0     \\
0          & K          & \Delta_{c0} & 0          & 0         & 0     \\
0          & -\zeta G^* & 0           & -\om_r     & -G        & 0     \\
-\zeta G^* & 0          & 0           & -G^*       & -\Delta_c & -K    \\
0          & 0          & 0           & 0          & -K^*      & - \Delta_{c0}
\end{pmatrix}
+\tilde D\,,
\end{align*}
where $\tilde D={\rm diag}[\gamma_m/2,\kappa_0,\kappa_{0f},\gamma_m/2,\kappa_0,\kappa_{0f}]$. Here, $K=\sqrt{N}h$ is the coupling of the center-of-mass mode to the bus cavity and the full OM coupling is again obtained for $\zeta=1$, while in RWA the off-diagonal blocks are dropped ($\zeta=0$).

We proceed to eliminate the OM degrees of freedom by applying the results of \appref{sec:appGeneralElimination} based on the interaction $H_{\rm int}=\lambda\sum_\nu(\ti\sigma^+_\nu \ti b_\nu+\hc)/2$ between qubits and resonators. The  $N$ sets of normal modes described by \eeqref{eq:QLEapp} are mutually uncorrelated, i.e., $\smean{\tilde b_i(t)\tilde b_j^\dagger(0)}_{\rm free}\propto \delta_{ij}$, and we may hence eliminate them one by one. Therefore, we drop the superfluous second index on the correlation functions \eqref{eq:corrS}-\eqref{eq:corrY} and finally note that the correlators for sets $i=2\ldots N$ are all identical, that is,  $\tilde X_{i\ge2}(\om)=\tilde X_2(\om)$, etc. The effective ME \eqref{eq:effME} may then be written in terms of collective and local contributions:
\begin{align}
\label{eq:MENnodes}
\dot \mu  &= -i\left[ J_N S^+S^-,\mu\right] -i \frac{\Delta}{2}\left[ S^z ,\mu\right]\\
&+\frac{1}{2}\Big\{ \Gamma_{\rm coll} (N_{\rm coll}+1) \mD[S^-]\mu 
+\Gamma_{\rm coll} N_{\rm coll} \mD[S^+]\mu \Big\} \nonumber\\
&+\frac{1}{2}\sum_{i=1}^N \Big\{ \Gamma_{\rm loc} (N_{\rm loc}+1) \mD[\sigma_i^-]\mu
+\Gamma_{\rm loc} N_{\rm loc} \mD[\sigma_i^+]\mu \Big\}\,.\nonumber
\end{align}
Here, we have introduced the collective operators $S^\pm=\sum_i \sigma_i^\pm$ and $S^z=\sum_i\sigma_z^i$, and the  interaction and decay rates are given by 
\begin{align}
J_N &= \frac{\lambda^2}{4N} \im\{ \tilde X_1 - \tilde X_2 \}\, , &
\Delta &=\frac{\lambda^2}{4} \im  \{ \tilde X_2 \}\,, \\
\Gamma_{\rm coll}&=\frac{\lambda^2}{2N}\re\{\tilde X_1-\tilde X_2\}\, , &
\Gamma_{\rm loc}&=\frac{\lambda^2}{2}\re\{\tilde X_2\}\,,
\end{align}
where we neglected a thermal contribution to $\Delta$. The collective and local occupation numbers are determined by
\begin{align}
\Gamma_{\rm coll}N_{\rm coll}&=\frac{\lambda^2}{4N}(\ti Y_1-\ti Y_2 )\, , \quad
\Gamma_{\rm loc}N_{\rm loc}=\frac{\lambda^2}{4}\ti Y_2\,,
\end{align}
and all correlation functions are understood to be evaluated at the qubit frequency $\om_q$. They can be evaluated as for the cascaded setting and with $\tilde A^i(\om)=(\tilde M^i-i\om\mathbb{1})^{-1}$ we obtain $\tilde X_i(\om) = \tilde A^i_{11}(\om)$ and $\ti Y_{i}(\om)= [\tilde A^{i*}(\om) \tilde r^i \tilde A^{iT}(\om)]_{11}$.

\subsection{Two node network}

For $N=2$ identical nodes the sets $\ti{\vecc{v}}^1$ and $\ti{\vecc{v}}^2$ correspond to the symmetric and anti-symmetric sets introduced in the main text as $\ti{\vecc{v}}^s$ and $\ti{\vecc{v}}^a$, respectively (cf. \eeqref{eq:QLEtwoNode}). In this case, the ME \eqref{eq:MENnodes} simplifies since the Lindblad terms are diagonal in the symmetric and anti-symmetric qubit operators $\ti\sigma_{s,a}^-=(\sigma_1^-\pm\sigma_2^-)/\sqrt{2}$:
\begin{align}
\label{eq:twoNodeMEonchip}
\dot\mu=
&-i \left[J(\sigma^+_1\sigma^-_2 + \sigma^-_1\sigma^+_2),\mu \right]
-i \frac{\Delta^\prime}{2} \sum_{i=1,2} \left[\sigma_z^i,\mu  \right] 
\\
&+\frac{1}{2}\sum_{\nu=a,s}\Big\{ \Gamma^\nu(N^\nu+1)\mD[\ti\sigma_\nu^-]\mu +\Gamma^\nu N^\nu\mD[\ti\sigma_\nu^+]\mu
\Big\}\, , \nonumber
\end{align}
where $\Delta^\prime=\Delta+J$ with $J=J_2$ and $\Delta$ given above, and 
\begin{align}
\Gamma^\nu &=\frac{\lambda^2}{2}\re\{ \tilde X_\nu\}\,,\,\quad
\Gamma^\nu N^\nu=\frac{\lambda^2}{4} \ti Y_\nu \,.
\end{align}
Finally, we estimate the fidelity for generating a Bell state as discussed in \secref{sec:onchipFidelity}. We add a term $\frac{1}{4T_2}\sum_i \mD[\sigma_z^i]$ to \eeqref{eq:twoNodeMEonchip} in order to account for intrinsic qubit dephasing and obtain to first order in the Lindblad terms:
\begin{align}
\label{eq:onchipFidelityFull}
\mF\approx 1 - \frac{\pi}{8}\sum_{\nu=s,a}\frac{\Gamma^\nu}{\abs{J}}(1+2N^\nu) - \frac{\pi}{8}\frac{1}{\abs{J} T_2}\,. 
\end{align}

%

\end{document}